\documentclass[letter]{IEEEtran}
\usepackage{amsmath,amsfonts}
\usepackage{algorithmic}
\usepackage{algorithm}
\usepackage{array}
\usepackage[caption=false,font=normalsize,labelfont=sf,textfont=sf]{subfig}
\usepackage{textcomp}
\usepackage{stfloats}
\usepackage{url}
\usepackage{verbatim}
\usepackage{graphicx}
\usepackage{cite}
\usepackage{epstopdf}
\usepackage{caption2}
\usepackage{diagbox}
\usepackage{subfig}
\usepackage{amssymb}
\usepackage{makecell}
\usepackage{mathrsfs}
\usepackage{bbding}
\usepackage{dsfont}
\usepackage{pifont}
\usepackage{xcolor}
\usepackage[colorlinks,
            allcolors=black,
            linkcolor=black,
            anchorcolor=black,
            citecolor=black]{hyperref}
\allowdisplaybreaks[4]

\begin{document}

\title{On the Secrecy Performance of Continuous-Aperture Arrays Over Fading Channels}

%\author{
%
%        % <-this % stops a space
%%\thanks{This paper was produced by the IEEE Publication Technology Group. They are in Piscataway, NJ.}% <-this % stops a space
%\thanks{This work was partially supported by National Natural Science Foundation of China Under Grants No. 62071250, the Fundamental Research Funds for the Central Universities under Grant 2242022R40030 and National Key Research and Development Project under Grant 2022YFB3904404. \emph{(Corresponding author: Dongming Li.)}}
%
%\thanks{X. Yang and D. Li are with School of Cyber Science and Engineering, Southeast University, Nanjing, 211189, China. (email:   xuan\_yang@seu.edu.cn; lidm@seu.edu.cn)}%
%}

\author{\IEEEauthorblockN{Xuan Yang,~\IEEEmembership{Graduate Student Member, IEEE}, Chongjun Ouyang,~\IEEEmembership{Member, IEEE}, Dongming Li,~\IEEEmembership{Member, IEEE}, Yuanwei Liu,~\IEEEmembership{Fellow, IEEE}}

\vspace{-1cm}

\thanks{X. Yang and D. Li are with School of Cyber Science and Engineering, Southeast University, Nanjing, 211189, China. (email: xuan\_yang@seu.edu.cn; lidm@seu.edu.cn)}\thanks{C. Ouyang is with the School of Electronic Engineering and Computer Science, Queen Mary University of London, London, E1 4NS, U.K.(email: c.ouyang@qmul.ac.uk)}\thanks{Y. Liu is with the Department of Electrical and Electronic Engineering, The University of Hong Kong, Hong Kong, China (email: yuanwei@hku.hk)}
}

% <-this % stops a space
%\thanks{This paper was produced by the IEEE Publication Technology Group. They are in Piscataway, NJ.}% <-this % stops a space

%\thanks{This work was partially supported by National Natural Science Foundation of China Under Grants No. 62071250, the Fundamental Research Funds for the Central Universities under Grant 2242022R40030 and National Key Research and Development Project under Grant 2022YFB3904404. \emph{(Corresponding author: Dongming Li.)}}
%
%\thanks{X. Yang and D. Li are with School of Cyber Science and Engineering, Southeast University, Nanjing, 211189, China. (email:   xuan\_yang@seu.edu.cn; lidm@seu.edu.cn)}%


%\IEEEpubidadjcol

\maketitle

\begin{abstract}
The secrecy performance of continuous-aperture array (CAPA)-based wiretap channels in terms of secrecy rate and secrecy outage probability (SOP) is analyzed. First, the system models of CAPA systems with maximum-ratio transmission under a Rayleigh fading channel are established, and approximate probability density functions for the legitimate user Bob's signal-to-noise ratio (SNR) and the eavesdropper Eve's SNR are derived using Mercer's theorem and Landau's eigenvalue theorem. Three scenarios are considered, including a single Eve, multiple independent Eves, and multiple collaborative Eves. Next, the expressions of the secrecy rate and SOP under these three scenarios are derived, and the high-SNR slope, high-SNR power offset, diversity order, and array gain in Bob's high-SNR region are obtained. It is then theoretically proven that, in all three scenarios, the CAPA system achieves the same high-SNR slope and the same diversity order, with the latter being equal to the spatial degrees of freedom. Moreover, the CAPA system with a single Eve has the smallest high-SNR offset and the highest array gain, whereas the CAPA system with multiple collaborative Eves exhibits the largest high-SNR offset and the lowest array gain. Finally, the theoretical analyses of secrecy rate, SOP, high-SNR performance are validated by the simulation results, and a higher secrecy rate and a lower SOP are achieved by the CAPA systems compared to the spatially-discrete array systems with half-wavelength antenna spacing.
\end{abstract}

\vspace{-0.3cm}
\section{Introduction}
Over the past several decades, multiple-input multiple-output (MIMO) technology has become a cornerstone of modern communication systems. By exploiting the advantages of multiple antennas, MIMO technology significantly improves system performance. The development trend of MIMO is to increase the number of antennas and operate at higher communication frequencies. Correspondingly, several new MIMO paradigms have emerged, such as gigantic MIMO \cite{li2023gigantic} and extremely large-scale MIMO \cite{wang2023extremely}. To further take the advantage of the ultimate performance of MIMO systems with an extremely large number of antennas within a given aperture area, the concept of continuous-aperture array (CAPA) \cite{liu2025capa} has been proposed in recent years. This concept is also referred to as holographic MIMO \cite{demir2022channel} or continuous-aperture MIMO (CAP-MIMO) \cite{wan2023can}. The concept of CAPA leverages virtually infinite radiating elements to realize arbitrary current distributions across the aperture. The shift from traditional spatially-discrete array (SPDA) to CAPA enables more flexible electromagnetic (EM) control, thereby allowing spatial multiplexing and diversity to be more fully exploited. Benefiting from advances in antenna technology, several CAPA system prototypes have already been developed to demonstrate proof of concept \cite{liu2025capa}.

Similar to the conventional SPDA systems, numerous studies have investigated the fundamental performance limits of CAPA systems. From an EM perspective, the Shannon information capacity of CAPA systems was analyzed in \cite{gruber2008new} based on Maxwell's equations. Based on this framework, the authors in \cite{wan2023mutual} employed Mercer expansion under a white-noise model and extended the analysis to the colored-noise scenario. The framework was later generalized to examine the uplink and downlink channel capacities of multi-user CAPA systems \cite{zhao2024continuous}. In addition to mutual information and channel capacity, the spatial degrees of freedom (DoF) of CAPA systems have attracted significant research attention \cite{miller2000communicating, ding2022degrees, pizzo2022landau, pizzo2022fourier}. Several methods have been proposed to analyze spatial DoF, including the eigenfunction approach \cite{miller2000communicating}, approximate analytical expressions \cite{dardari2020communicating}, Fourier plane-wave series expansion \cite{pizzo2020spatially}, and Landau's eigenvalue theorem-based approach \cite{pizzo2022landau}. Other works have examined the impact of specific factors on spatial DoF, such as coupling gain \cite{dardari2020communicating}, geometric configurations \cite{ding2022degrees}, and near-field communication \cite{pizzo2022fourier}. Since diversity and multiplexing are two key metrics in wireless systems, related studies have also investigated their behavior in CAPA systems. In particular, the diversity and multiplexing performance of two non-parallel continuous-aperture arrays (CAPAs) was analyzed in \cite{ouyang2025diversity}, where the diversity-multiplexing trade-off (DMT) was also characterized.

Beyond analyzing the fundamental performance limits of CAPA systems, beamforming design in CAPA systems has also attracted significant research interest. Beamforming design, which is also referred to as current distribution or pattern design in CAPA systems, is a key factor in determining the communication performance \cite{dardari2020communicating}. In the CAPA systems, square-wave functions \cite{dardari2020communicating} and Fourier basis functions \cite{sanguinetti2022wavenumber} have been employed within the wavenumber-division multiplexing (WDM) scheme for downlink beamforming design. The WDM scheme has further been extended to planar CAPA systems for both downlink \cite{zhang2023pattern} and uplink \cite{qian2024spectral} multi-user channels, where optimal solutions are obtained under sum-rate maximization and spectral efficiency maximization criteria, respectively. To reduce the computational complexity of Fourier-based approaches \cite{zhang2023pattern, qian2024spectral}, optimal beamforming methods based on the calculus of variations (CoV) have been proposed \cite{wang2025optimal, wang2025beamforming}. Beyond transmitter-side beamforming design \cite{qian2024spectral}, receiver beamforming schemes under various optimization criteria have also been developed \cite{ouyang2025linear}. In addition, deep learning has been demonstrated as an effective tool for beamforming design in CAPA systems \cite{guo2024deep}.

Although there is a growing body of work on beamforming design in CAPA systems, most studies primarily focus on system effectiveness and reliability. The security performance, which constitutes the third key characteristic of wireless systems, has been seldom investigated, and only a few works have addressed this aspect \cite{zhao2024physical, sun2025secure}. In \cite{zhao2024physical}, the authors considered a single-user and single-eavesdropper scenario under a line-of-sight (LoS) channel and derived the optimal current distributions to maximize the secrecy rate under a power constraint, as well as the minimum required power to achieve a target secrecy rate. Compared to the SPDA systems, the superiority of CAPA systems in terms of physical-layer security (PLS) performance was demonstrated. This system model was later extended to the multi-user and multi-eavesdropper scenarios in \cite{sun2025secure}. Under the criterion of maximizing the weighted secrecy sum-rate, secure beamforming was designed using a block coordinate descent optimization method and a zero-forcing-based heuristic algorithm. Nevertheless, both \cite{zhao2024physical} and \cite{sun2025secure} consider LoS channels. The key PLS performance metrics, such as secrecy rate, secrecy outage probability (SOP), and high-signal-to-noise ratio (high-SNR) performance under non-line-of-sight (NLoS) channels in CAPA systems, warrant further investigation.

To address this gap, we investigate the secrecy rate, SOP, and high-SNR secrecy performance of the CAPA systems under Rayleigh fading channels, explicitly accounting for channel correlation. To the best of our knowledge, this is the first work to analyze the security performance of the CAPA systems under Rayleigh fading channels while incorporating the effect of channel correlation. The main contributions of this paper are summarized as follows:

\hangindent=2em
\hangafter=0
\noindent\textbullet\ We construct the channel model in CAPA systems using a Gaussian random field and establish system models for CAPA with maximum-ratio transmission (MRT) under Rayleigh fading, considering scenarios with two legitimate users under a single eavesdropper (Eve), multiple independent eavesdroppers, and multiple collaborative eavesdroppers, respectively. We then obtain the channel autocorrelation function through distributional equivalence and decompose it using Mercer's theorem and Landau's eigenvalue theorem to derive the corresponding eigenvalues and eigenfunctions. Moreover, based on these decomposed eigenvalues and eigenfunctions, we derive approximate probability density functions (PDFs) for the legitimate user Bob's signal-to-noise ratio (SNR) and Eve's SNR in the three scenarios.

\begin{figure}[htbp]%
    \centering
    \subfloat{
        \includegraphics[width=7cm]{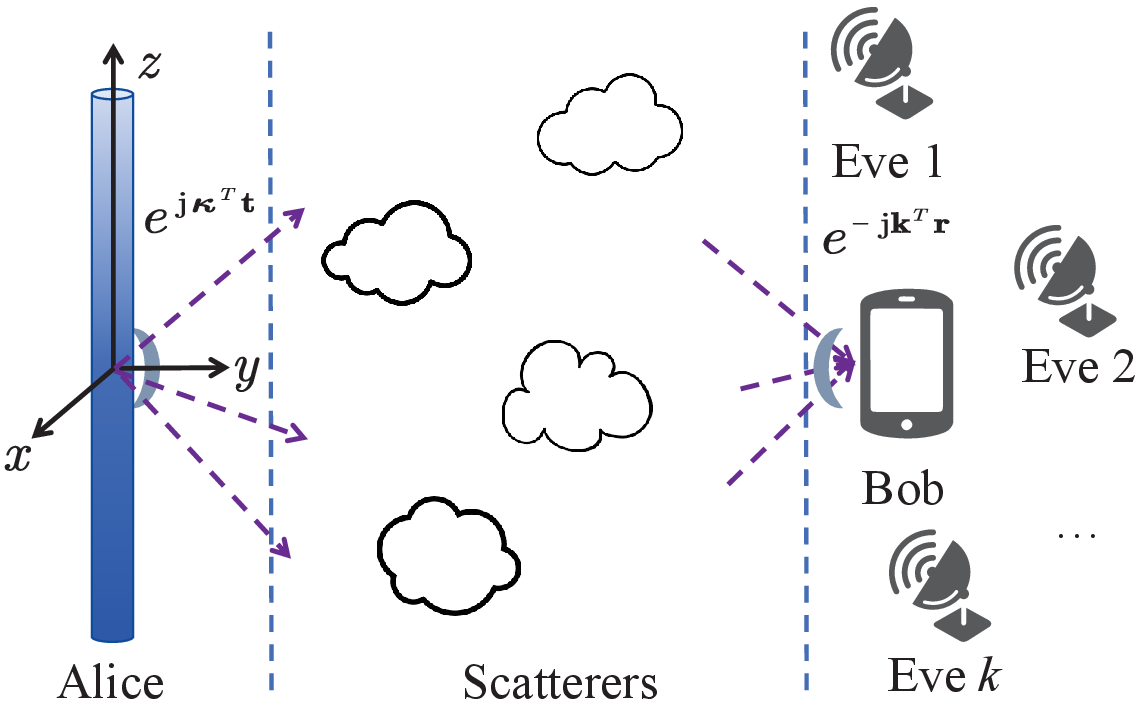}
        }
    \caption{CAPA system model with eavesdropping.}
    \label{fig1}
    \vspace{-0.4cm}
\end{figure}

\hangindent=2em
\hangafter=0
\noindent\textbullet\ We derive analytical expressions for the secrecy rate and SOP under the aforementioned eavesdropping scenarios. Furthermore, we analyze the high-SNR slope, high-SNR power offset, diversity order, and array gain in Bob's high-SNR regime. It is shown that, across all three scenarios, the high-SNR slope remains identical and the diversity order in each case equals the system's spatial degrees of freedom (DoF). Meanwhile, we find that the CAPA system with a single Eve achieves the smallest high-SNR offset and the highest array gain, whereas the CAPA system with multiple collaborative Eves exhibits the largest high-SNR offset and the lowest array gain.

\hangindent=2em
\hangafter=0
\noindent\textbullet\ We validate the theoretical analyses through simulations, covering secrecy rate, SOP, and high-SNR secrecy performance of the CAPA systems. The simulation results show that: i) The CAPA systems achieve a higher secrecy rate and a lower SOP than the SPDA systems under half-wavelength aperture spacing; ii) The CAPA system achieves the same high-SNR slope and the same diversity order, with the diversity order being equal to the spatial degrees of freedom; iii) The CAPA system with a single Eve demonstrates the smallest high-SNR offset and the highest array gain, whereas the CAPA system with multiple collaborative Eves exhibits the largest high-SNR offset and the lowest array gain.

The remainder of this paper is organized as follows: Section II introduces the system model and the statistical characteristics of the CAPA channel. In Section III, we derive PDFs of Bob's and Eve's signal-to-noise ratios (SNRs) under different scenarios. Secrecy rate and SOP are analyzed in Section IV (a single Eve), Section V (multiple independent Eves), and Section VI (multiple collaborative Eves). Section VII presents simulation results to validate the theoretical analyses and demonstrate the superiority of the CAPA systems over the SPDA systems. Finally, Section VIII concludes the paper.

\textit{Notations}: Scalars and vectors are denoted by non-bold and bold lowercase letters, respectively. The operators $\left( \cdot \right)^{\mathrm{T}}$ and $\left( \cdot \right)^*$ denote the transpose and conjugate operations, respectively. The expectation operator is represented by $\mathbb{E} \left\{ \cdot \right\} $. The natural logarithm is denoted by $\ln(\cdot)$, and $\overset{d}{=}$ indicates equality in distribution. The magnitude of a scalar $a$ is denoted by $\left| a \right|$. The Kronecker delta function is represented by $\delta_{i,j}$, and the Dirac delta function by $\delta(\cdot)$. Finally, $\mathbb{R}$ and $\mathbb{Z}^+$ denote the set of real numbers and the set of positive integers, respectively.

\vspace{-0.4cm}
\section{System Model}
\subsection{Signal Model Under MRT}

As shown in Fig. \ref{fig1}, the system model considered in this work consists of two legitimate users, Alice and Bob, and either a single Eve or multiple Eves, where Alice is equipped with a linear CAPA and both Bob and Eves are equipped with a single antenna. Alice, Bob, and Eve are denoted by $\mathrm{a}$, $\mathrm{b}$, and $\mathrm{e}$, respectively. There is no LOS path between Alice and Bob, nor between Alice and Eves. The CAPA, with length $L$, is aligned along the $z$-axis and symmetric with respect to the $x$-$y$ plane. The transmitted signal from Alice can be expressed as follows:
\begin{equation}
\label{eq1}
x\left( \mathbf{t} \right) =\sqrt{P}w\left( \mathbf{t} \right) s,\mathbf{t}\in \mathcal{A},
\end{equation}
where $s$ is the normalized coded data symbol satisfying the power constraint $\mathbb{E} \left\{ \left| s \right|^2 \right\} =1$, $P$ denotes the transmission power, ${{w}}(\mathbf{t})$ is the precoding vector, and $\mathcal{A} =\left\{ \left[ 0,0,z \right] ^{\mathrm{T}}|z\in \left[ -\frac{L}{2},\frac{L}{2} \right] \right\} $ represents the aperture of CAPA. After transmission through a frequency-flat Rayleigh fading channel, the signal received by Bob is given by
\begin{equation}
\label{eq2}
y_{\mathrm{b}}\left( \mathbf{r} \right) =\sqrt{P}s\int_{\mathcal{A}}{h_{\mathrm{b}}\left( \mathbf{r},\mathbf{t} \right) w\left( \mathbf{t} \right) d\mathbf{t}}+n_{\mathrm{b}}\left( \mathbf{r} \right) ,
\end{equation}
where $\mathbf{r}$ denotes the location of Bob, $h_\mathrm{b}(\mathbf{r},\mathbf{t})$ the channel between Bob and the point $\mathbf{t}$ on CAPA. The term $n_\mathrm{b}(\mathbf{r})$ denotes complex Gaussian noise with zero mean and unit variance.

Let $K$ denote the number of Eves. For each $k\in \left\{ 1,\cdots ,K \right\} $, the signal received by Eve $k$ is given by
\begin{equation}
\label{eq3}
y_{\mathrm{e}_k}\left( \mathbf{r} \right) =\sqrt{P}s\int_{\mathcal{A}}{h_{\mathrm{e}_k}\left( \mathbf{r},\mathbf{t} \right) w\left( \mathbf{t} \right) d\mathbf{t}}+n_{\mathrm{e}_k}\left( \mathbf{r} \right) ,
\end{equation}
where $h_{\mathrm{e}_k}(\mathbf{r},\mathbf{t})$ denotes the channel between Eve $k$ and the point $\mathbf{t}$ on CAPA. It is assumed that $h_{\mathrm{e}_k}(\mathbf{r},\mathbf{t})$ is a zero-mean, unit-variance complex Gaussian (ZUCG) random process, and is independent of $h_\mathrm{b}(\mathbf{r},\mathbf{t})$. The term $n_{\mathrm{e}_k}(\mathbf{r})$ represents complex Gaussian noise with zero mean and unit variance.

To maximize the received signal power at Bob, MRT is applied at Alice. Accordingly, we have
\begin{equation}
\label{eq4}
w\left( \mathbf{t} \right) =\frac{h_{\mathrm{b}}^{*}\left( \mathbf{r},\mathbf{t} \right)}{\sqrt{\int_{\mathcal{A}}{\left| h_{\mathrm{b}}\left( \mathbf{r},\mathbf{t} \right) \right|^2d\mathbf{t}}}}.
\end{equation}

It follows that
\begin{equation}
\label{eq5}
\begin{aligned}
&y_{\mathrm{b}}\left( \mathbf{r} \right) =\sqrt{P}s\cdot \sqrt{\int_{\mathcal{A}}{\left| h_{\mathrm{b}}\left( \mathbf{r},\mathbf{t} \right) \right|^2dt}}+n_{\mathrm{b}}\left( \mathbf{r} \right) ,
\\
&y_{\mathrm{e}_k}\left( \mathbf{r} \right) =\sqrt{P}s\cdot \frac{\int_{\mathcal{A}}{h_{\mathrm{e}_k}\left( \mathbf{r},\mathbf{t} \right) h_{\mathrm{b}}^{*}\left( \mathbf{r},\mathbf{t} \right) d\mathbf{t}}}{\sqrt{\int_{\mathcal{A}}{\left| h_{\mathrm{b}}\left( \mathbf{r},\mathbf{t} \right) \right|^2d\mathbf{t}}}}+n_{\mathrm{e}_k}\left( \mathbf{r} \right) .
\end{aligned}
\end{equation}

\subsection{Channel Model}
The scatterers are assumed to be confined to the region between Alice, Bob, and Eves. Under this assumption, the EM multipath spatial response $h_\mathrm{b}(\mathbf{r},\mathbf{t})$ between Alice and Bob can be modeled as \cite{pizzo2020spatially, pizzo2022spatial}
\begin{equation}
\label{eq6}
h_{\mathrm{b}}(\mathbf{r},\mathbf{t})=\iiiint_{\mathscr{D} \left( \boldsymbol{\kappa } \right) \times \mathscr{D} (\mathbf{k})}{\frac{e^{-\mathrm{j}\mathbf{k}^{\mathrm{T}}\mathbf{r}}H_{\mathrm{b}}(\mathbf{k},\boldsymbol{\kappa })e^{\mathrm{j}\boldsymbol{\kappa }^{\mathrm{T}}\mathbf{t}}}{(2\pi )^2}\mathrm{d}\mathbf{k}\mathrm{d}\boldsymbol{\kappa },}
\end{equation}
where $\boldsymbol{\kappa }=\left[ \kappa _x,\gamma \left( \kappa _x,\kappa _z \right) ,\kappa _z \right] ^{\mathrm{T}}\in \mathbb{R} ^{3\times 1}$, $\mathbf{k}=\left[ k_x,\gamma \left( k_x,k_z \right) ,k_z \right] ^{\mathrm{T}}\in \mathbb{R} ^{3\times 1}$, $\gamma (x,y)=\left( k_{0}^{2}-x^2-y^2 \right) ^{\frac{1}{2}}$, $\mathscr{D} (\boldsymbol{\kappa })=\left\{ \left( \kappa _x,\kappa _z \right) \in \mathbb{R} ^2\mid \kappa _{x}^{2}+\kappa _{z}^{2}\le k_{0}^{2} \right\}$, and $\mathscr{D} (\mathbf{k})=\left\{ \left( k_x,k_z \right) \in \mathbb{R} ^2\mid k_{x}^{2}+k_{z}^{2}\le k_{0}^{2} \right\} $. Here, $k_0$ is the wavenumber, given by $k_0 = \frac{2\pi}{\lambda}$, where $\lambda$ is the wavelength. The terms $e^{j \boldsymbol{\kappa}^{\mathrm{T}} \mathbf{t}}$ and $e^{j \mathbf{k}^{\mathrm{T}} \mathbf{t}}$ represent the transmit and receive plane waves in the directions of $\frac{\boldsymbol{\kappa}}{|\boldsymbol{\kappa}|}$ and $\frac{\mathbf{k}}{|\mathbf{k}|}$, respectively. Moreover, $H_\mathrm{b}(\mathbf{k}, \boldsymbol{\kappa})$ denotes the angular response that maps the transmit direction $\frac{\boldsymbol{\kappa}}{|\boldsymbol{\kappa}|}$ to the receive direction $\frac{\mathbf{k}}{|\mathbf{k}|}$, and is modeled as a random field. For analytical tractability, we assume Rayleigh fading and isotropic scattering \cite{pizzo2020spatially, pizzo2022spatial}, yielding
\begin{equation}
\label{eq7}
H_\mathrm{b}(\mathbf{k},\boldsymbol{\kappa })=\frac{\sigma _{h_\mathrm{b}}A_S\left( k_0 \right)}{\sqrt{\gamma \left( k_x,k_z \right) \gamma \left( \kappa _x,\kappa _z \right)}}W_\mathrm{b}(\mathbf{k},\boldsymbol{\kappa }),
\end{equation}
where $W_\mathrm{b}(\mathbf{k}, \boldsymbol{\kappa})$ denotes a ZUCG random field defined over $\mathscr{D}(\boldsymbol{\kappa}) \times \mathscr{D}(\mathbf{k})$. The normalization constant $A_S(k_0)$ is set to $\frac{2\pi}{k_0}$ to ensure that the channel gain satisfies $\sigma_{h_\mathrm{b}}^2$, i.e., $\mathbb{E} \{\left| h_\mathrm{b}(\mathbf{r},\mathbf{t}) \right|^2\}=\sigma _{h_\mathrm{b}}^{2}$.

Using a similar approach, the EM multipath spatial response $h_{\mathrm{e}_k}(\mathbf{r}, \mathbf{t})$ between Alice and Eve $k$ can be expressed as follows:
\begin{equation}
\label{eq8}
h_{\mathrm{e}_k}(\mathbf{r},\mathbf{t})=\iiiint_{\mathscr{D} \left( \boldsymbol{\kappa } \right) \times \mathscr{D} (\mathbf{k})}{\frac{e^{-\mathrm{j}\mathbf{k}^{\mathrm{T}}\mathbf{r}}H_{\mathrm{e}_k}(\mathbf{k},\boldsymbol{\kappa })e^{\mathrm{j}\boldsymbol{\kappa }^{\mathrm{T}}\mathbf{t}}}{(2\pi )^2}\mathrm{d}\mathbf{k}\mathrm{d}\boldsymbol{\kappa },}
\end{equation}
where
\begin{equation}
\label{eq9}
H_{\mathrm{e}_k}(\mathbf{k},\boldsymbol{\kappa })=\frac{\sigma _{h_{\mathrm{e}_k}}A_S\left( k_0 \right)}{\sqrt{\gamma \left( k_x,k_z \right) \gamma \left( \kappa _x,\kappa _z \right)}}W_{\mathrm{e}_k}(\mathbf{k},\boldsymbol{\kappa }),
\end{equation}
and where $W_{\mathrm{e}_k}(\mathbf{k}, \boldsymbol{\kappa})$ denotes a ZUCG random field defined over $\mathscr{D}(\boldsymbol{\kappa}) \times \mathscr{D}(\mathbf{k})$.

\subsection{Channel Statistics}
\subsubsection{Statistical Equivalence}
By substituting $H_\mathrm{b}(\mathbf{k},\boldsymbol{\kappa })$ from (\ref{eq7}) into (\ref{eq6}), we obtain
\begin{equation}
\label{eq10}
h_{\mathrm{b}}(\mathbf{r},\mathbf{t})=\frac{\sigma _{h_{\mathrm{b}}}}{(2\pi )^2}\iint_{\mathscr{D} (\boldsymbol{\kappa })}{\frac{e^{\mathrm{j}\boldsymbol{\kappa }^{\mathrm{T}}\mathbf{t}}}{\sqrt{\gamma (\kappa _x,\kappa _z)}}\hat{H}_{\mathrm{b}}(\boldsymbol{\kappa })\mathrm{d}\boldsymbol{\kappa },}
\end{equation}
where $\hat{H}_{\mathrm{b}}(\boldsymbol{\kappa })=\iint_{\mathscr{D} (\mathbf{k})}{\frac{A_S(k_0)e^{-\mathrm{j}\mathbf{k}^{\mathrm{T}}\mathbf{t}}}{\sqrt{\gamma (k_x,k_z)}}W_{\mathrm{b}}(\mathbf{k},\boldsymbol{\kappa })\mathrm{d}\mathbf{k}}$. Since $W_\mathrm{b}(\mathbf{k},\boldsymbol{\kappa })$ is a ZUCG random field, $\hat{H}_\mathrm{b}(\boldsymbol{\kappa })$ is a zero-mean complex Gaussian (ZMCG) random field. With basic mathematical manipulations, it can be shown that $\hat{H}_\mathrm{b}(\boldsymbol{\kappa })\overset{d}{=}\hat{W}_\mathrm{b}(\boldsymbol{\kappa })A_S(k_0)(2\pi k_0)^{\frac{1}{2}}$, where $\hat{W}_\mathrm{b}(\boldsymbol{\kappa })$ denotes a ZUCG random field defined over $\mathscr{D} (\mathbf{k})$.

Since $\boldsymbol{\kappa }=[\kappa _x,\gamma (\kappa _x,\kappa _z),\kappa _z]^{\mathrm{T}}$ and $\mathbf{t}=[0,0,z]^{\mathrm{T}}$ for $z\in \left[ -\frac{L}{2},\frac{L}{2} \right]$, (\ref{eq10}) can be simplified as follows:
\begin{equation}
\label{eq11}
h_{\mathrm{b}}(\mathbf{r},\mathbf{t})=\frac{\sigma _{h_{\mathrm{b}}}}{(2\pi )^2}\int_{-k_0}^{k_0}{e^{\mathrm{j}\kappa _zz}H_{\mathrm{b}}(\kappa _z)\mathrm{d}\kappa _z}\triangleq g_{\mathrm{b}}(z),
\end{equation}
where $H_\mathrm{b}(\kappa _z)=\int_{-\sqrt{k_{0}^{2}-\kappa _{z}^{2}}}^{\sqrt{k_{0}^{2}-\kappa _{z}^{2}}}{\frac{\hat{H}_\mathrm{b}(\mathbf{\boldsymbol{\kappa}})}{\sqrt{\gamma (\kappa _x,\kappa _z)}}\mathrm{d}\kappa _x}$ is a ZMCG random filed defined over $\left[ -k_0,k_0 \right]$. Since $\hat{H}_\mathrm{b}(\boldsymbol{\kappa })\overset{d}{=}\hat{W}_\mathrm{b}(\boldsymbol{\kappa })A_S(k_0)(2\pi k_0)^{\frac{1}{2}}$, it follows that
\begin{equation}
\label{eq12}
H_\mathrm{b}(\kappa _z)\overset{d}{=}A_S(k_0)(2\pi k_0)^{\frac{1}{2}}\pi ^{\frac{1}{2}}W_\mathrm{b}(\kappa _z),
\end{equation}
where $W_\mathrm{b}(\kappa _z)$ denotes a ZUCG random field defined over $\left[ -k_0,k_0 \right] $. It then follows that $g_\mathrm{b}\left( z \right)$ is a ZMCG random field defined over $\left[ -\frac{L}{2},\frac{L}{2} \right]$. Consequently, the statistics of $g_\mathrm{b}\left( z \right)$ are fully characterized by its autocorrelation function, which can be expressed as follows:
\begin{equation}
\label{eq13}
\begin{aligned}
R_{g_\mathrm{b}}(z,z^{\prime})=&\mathbb{E} \{g_\mathrm{b}(z)g_\mathrm{b}^{*}(z^{\prime})\}=\frac{\sigma _{h_\mathrm{b}}^{2}}{(2\pi )^4}\int_{-k_0}^{k_0}{\int_{-k_0}^{k_0}{}}
\\
&\mathbb{E} \{H_\mathrm{b}(\kappa _z)H_\mathrm{b}^{*}(\kappa _{z}^{\prime})\}\mathrm{e}^{\mathrm{j}\kappa _zz}e^{-\mathrm{j}\kappa _{z}^{\prime}z^{\prime}}\mathrm{d}\kappa _z\mathrm{d}\kappa _{z}^{\prime}.
\end{aligned}
\end{equation}

\subsubsection{Autocorrelation}
Base on (\ref{eq12}), it follows that $\mathbb{E} \{H_\mathrm{b}(\kappa _z)H_\mathrm{b}^{*}(\kappa _{z}^{\prime})\}=A_{S}^{2}(k_0)(2\pi k_0)\pi \delta (\kappa _z-\kappa _{z}^{\prime})$. Then, it follows that
\begin{equation}
\label{eq14}
R_{g_{\mathrm{b}}}(z,z^{\prime})=\frac{\sigma _{h_{\mathrm{b}}}^{2}}{2k_0}\int_{-k_0}^{k_0}{e^{\mathrm{j(}z-z^{\prime})\kappa _z}\mathrm{d}\kappa _z}=\frac{\sigma _{h_\mathrm{b}}^{2}\sin \left[ k_0\left( z-z^{\prime} \right) \right]}{k_0\left( z-z^{\prime} \right)}.
\end{equation}
Thus, the autocorrelation function $R_{g_\mathrm{b}}(z,z^{\prime})$ is a bounded, symmetric, and positive semidefinite Hilbert-Schmidt operator \cite{ghanem2003stochastic}. According to Mercer's theorem \cite{hilbert1985methods}, by performing spectral decomposition \cite{ghanem2003stochastic} on $R_{g_\mathrm{b}}(z,z^{\prime})$, we obtain:
\begin{equation}
\label{eq15}
R_{g_{\mathrm{b}}}(z,z^{\prime})=\sum_{\ell =1}^{\infty}{\sigma _{h_{\mathrm{b}}}^{2}\sigma _{\ell}\phi _{\ell}(z)\phi _{\ell}^{*}(z^{\prime})}\approx \sum_{\ell =1}^T{\sigma _{h_{\mathrm{b}}}^{2}\sigma _{\ell}\phi _{\ell}(z)\phi _{\ell}^{*}(z^{\prime}),}
\end{equation}
where $\sigma _{h_\mathrm{b}}^{2}\sigma _{\ell}$, $\ell=1,2,...,\infty$, are the eigenvalues of $R_{g_\mathrm{b}}(z,z^{\prime})$, satisfying $\sigma _1\geqslant \sigma _2\geqslant ...\geqslant \sigma _{\infty}\geqslant 0$. The corresponding eigenfunctions $\phi _{\ell}\left( \cdot \right) ,\ell=1,2,...,\infty $, form an orthogonal basis over $\left[ -\frac{L}{2},\frac{L}{2} \right]$, such that $\int_{-\frac{L}{2}}^{\frac{L}{2}}{\phi _{\ell}^{*}(z)\phi _{\ell ^{\prime}}(z)\mathrm{d}z}=\delta _{\ell ,\ell ^{\prime}}$, where $\delta _{\ell ,\ell ^{\prime}}$ is the Kronecker delta function. The first $T$ eigenfunctions provide an accurate approximation of $R_{g_{\mathrm{b}}}(z,z^{\prime})$ \cite{hilbert1985methods}. The eigenvalues $\sigma _{h_\mathrm{b}}^{2}\sigma _{\ell}$ can be obtained using Guass-Legendre quadrature \cite{swarztrauber2003computing} in combination with singular value decomposition (SVD). For further details, please refer to Appendix A.

Next, we exploited additional properties of the eigenvalues to gain deeper insight. Define $K(z,z^{\prime})\triangleq \tfrac{1}{2\pi}\int_{-k_0}^{k_0}{{e}^{\mathrm{j(}z-z^{\prime})\kappa _z}\mathrm{d}\kappa _z}$, where $z,z^{\prime}\in \left[ -\frac{L}{2},\frac{L}{2} \right]$. It then follows that $R_{g_{\mathrm{b}}}(z,z^{\prime})=\tfrac{2\pi}{2k_0}\sigma _{h_{\mathrm{b}}}^{2}K(z,z^{\prime})$. The eigenvalues of $K(z,z^{\prime})$ are denoted by $\varepsilon _\ell$, $\ell=1,2,...,\infty $, satisfying $\varepsilon _1\geqslant \varepsilon _2\geqslant ...\geqslant \varepsilon _{\infty}\geqslant 0$. Consequently, the eigenvalues of $R_{g_\mathrm{b}}(z,z^{\prime})$ are given by $\sigma _{\ell}=\tfrac{2\pi \varepsilon _{\ell}}{2k_0}=\frac{\lambda}{2}\varepsilon _{\ell}$.

Assume that $f\left( z^{\prime} \right)$ is a square-integrable function defined over $\left[ -\frac{L}{2},\frac{L}{2} \right] $. We define $\hat{f}\left( z \right) \triangleq \int_{-\frac{L}{2}}^{\frac{L}{2}}{K\left( z,z^{\prime} \right) f\left( z^{\prime} \right) dz^{\prime}}$. Then, $\hat{f}\left( z \right)$ can be rewritten as follows:
\begin{equation}
\label{eq16}
\hat{f}\left( z \right) =\mathds1_{\left[ -\frac{L}{2},\frac{L}{2} \right]}\left( z \right) \int_{-\frac{L}{2}}^{\frac{L}{2}}{k\left( z-z^{\prime} \right) f\left( z^{\prime} \right) dz^{\prime},}
\end{equation}
where $\mathds{1} _{\left[ \cdot ,\cdot \right]}\left( \cdot \right)$ denotes the indicator function, defined as follows:
\begin{equation}
\label{eq17}
\mathds{1} _{\left[ a,b \right]}\left( z \right) =\left\{ \begin{array}{c}
	1,z\in \left[ a,b \right]\\
	0,\mathrm{others}.\\
\end{array} \right.
\end{equation}
Meanwhile, $k\left( \cdot \right) $ is defined as $k(x)\triangleq \tfrac{1}{2\pi}\int_{-k_0}^{k_0}{{e}^{\mathrm{j}x\kappa _z}\mathrm{d}\kappa _z},x\in \mathbb{R}$. Therefore, $k\left( \cdot \right)$ is the inverse Fourier transform of $\mathds{1} _{\left[ -k_0,k_0 \right]}\left( \kappa _z \right)$, and $k\left( \cdot \right)$ can be interpreted as an ideal filter over the frequency range $\left[ -k_0,k_0 \right]$ after Fourier transformation. Within this framework, Landau's eigenvalue theorem \cite{landau1980eigenvalue} can be applied, and the eigenvalues $\left\{ \varepsilon _\ell \right\} _{\ell=1}^{\infty}$ can be characterized as follows:
\begin{equation}
\label{eq18}
1\ge \varepsilon _1\ge \varepsilon _2...\ge \varepsilon _{\infty}\ge 0.
\end{equation}
For $\varepsilon >0$ and $L\gg \lambda$, we obtain:
\begin{equation}
\label{eq19}
|\{\ell :\varepsilon _{\ell}>\varepsilon \}|=\mathsf{DOF}
\\
+\left( \frac{1}{\pi ^2}\ln \frac{1-\sqrt{\varepsilon}}{\sqrt{\varepsilon}} \right) \ln \mathsf{DOF}+o(\ln \mathsf{DOF}),
\end{equation}
where $\mathsf{DOF}=\frac{2L}{\lambda}$ denotes the effective spatial DoF of the CAPA system  \cite{ouyang2025diversity}.

{\bf{Remark 1.}} {\it{As indicated by (\ref{eq18}) and (\ref{eq19}), when $L\gg \lambda$, the eigenvalues $\left\{ \varepsilon _\ell \right\} _{\ell=1}^{\infty}$ remain close to one before rapidly decaying to zero. The width of this transition region is proportional to $\ln\mathsf{DOF}$. Owing to this rapid decay, the first $\mathsf{DOF}$ eigenvalues can be used to approximate $R_{g_\mathrm{b}}(z,z^{\prime})$ \cite{ouyang2025diversity}.}}

\vspace{-0.3cm}
\section{PDFs of Bob's and Eve's SNRs under Different Scenarios}
In this section, we derive PDFs of Bob's and Eve's SNRs under MRT beamforming. Three scenarios are considered: a single Eve, multiple independent Eves, and multiple collaborative Eves.
\subsection{PDF of Bob's SNR}
The statistics of $g_\mathrm{b}\left( z \right)$ are entirely determined by its autocorrelation function $R_{g_\mathrm{b}}\left( z,z^{\prime} \right)$, since $g_\mathrm{b}\left( z \right)$ is a ZMCG random field. From (14), it follows that
\begin{equation}
\label{eq20}
g_{\mathrm{b}}(z)\overset{d}{=}\overline{g}_{\mathrm{b}}(z)=\sigma _{h_{\mathrm{b}}}\sum_{\ell =1}^{\infty}{\int_{-\frac{L}{2}}^{\frac{L}{2}}{\sigma _{\ell}^{\frac{1}{2}}\phi _{\ell}(z)\phi _{\ell}^{*}(z^{\prime})\overline{W}_{\mathrm{b}}(z^{\prime})\mathrm{d}z^{\prime},}}
\end{equation}
where $\overline{W}_\mathrm{b}(z)$ is a ZUCG random field defined over $\left[ -\frac{L}{2},\frac{L}{2} \right]$. It can be verified that $g_{\mathrm{b}}(z)$ and $\overline{g}_{\mathrm{b}}(z)$ share the same autocorrelation function. Consequently, $g_{\mathrm{b}}(z)$ and $\overline{g}_{\mathrm{b}}(z)$ have identical statistical characteristics, as both are ZMCG random fields. It then follows that the instantaneous SNR of Bob is
\begin{equation}
\label{eq21}
\rho _{\mathrm{b}}=\bar{\rho}_{\mathrm{b}}\int_{\mathcal{A}}{\left| h_{\mathrm{b}}(\mathbf{r},\mathbf{t}) \right|^2\mathrm{d}\mathbf{t}}\mathop {=}^d\bar{\rho}_{\mathrm{b}}\int_{-\frac{L}{2}}^{\frac{L}{2}}{\left| \bar{g}_{\mathrm{b}}(z) \right|^2\mathrm{d}z,}
\end{equation}
where ${\bar \rho _{\mathrm{b}}} = \frac{{{P}}}{{\sigma _{\mathrm{b}}^2}}$, and $P$ denotes the transmit power of $x\left( \mathbf{t} \right)$. We define $\Phi _{\mathrm{b},\ell}\overset{d}{=}\int_{-\frac{L}{2}}^{\frac{L}{2}}{\phi _{\ell}^{*}(z^{\prime})\overline{W}_{\mathrm{b}}(z^{\prime})\mathrm{d}z^{\prime}}$. Since $\overline{W}_\mathrm{b}(z)$ is a ZUCG random field, $\Phi _{\mathrm{b},\ell}$ is a complex Gaussian random variable. Moreover, due to the orthogonality of $\{\phi _{\ell}(\cdot )\}_{\ell =1}^{\infty}$, the set $\{\Phi _{\mathrm{b},\ell}\}_{\ell =1}^{\infty}$ consists of independently and identically distributed (i.i.d.) complex Gaussian random variables with zero mean and unit variance. Thus, we can get
\begin{equation}
\label{eq22}
\rho _{\mathrm{b}}\overset{d}{=}\overline{\gamma }_{\mathrm{b}}\sum_{\ell =1}^{\infty}{\sigma _{\ell}|\Phi _{\mathrm{b},\ell}|^2,}
\end{equation}
where $\overline{\gamma }_\mathrm{b}$ denotes the average SNR of Bob, given by $\overline{\gamma }_\mathrm{b}=\frac{P\sigma _{h_\mathrm{b}}^{2}}{\sigma _\mathrm{b}^{2}}$. As noted in Remark 1, for $i=1,2,\cdots ,\mathsf{DOF}$, the eigenvalues satisfy $\varepsilon _i\approx 1$ and $\sigma _i\approx \frac{\lambda}{2}$. For $i>\mathsf{DOF}$, both $\varepsilon _i$ and $\sigma _i$ rapidly decrease and approach zero. This step-like behavior becomes more pronounced as $L\gg \lambda $. Consequently, the instantaneous SNR of Bob's received signal can be expressed as follows:
\begin{equation}
\label{eq23}
\rho _{\mathrm{b}}\overset{d}{=}\overline{\gamma }_{\mathrm{b}}\sum_{\ell =1}^{\infty}{\sigma _{\ell}|\Phi _{\mathrm{b},\ell}|^2}\approx \overline{\gamma }_{\mathrm{b}}\sum_{\ell =1}^{\mathsf{DOF}}{\sigma _{\ell}|\Phi _{\mathrm{b},\ell}|^2.}
\end{equation}
Thus, $\rho _\mathrm{b}$ can be asymptotically approximated as the sum of independent exponentially distributed random variables with different weights. The PDF of $\rho _\mathrm{b}$ \cite{moschopoulos1985distribution} can then be expressed as
\begin{equation}
\label{eq24}
f_{\rho _{\mathrm{b}}}(x)\approx \frac{1}{\overline{\gamma }_{\mathrm{b}}\prod_{\ell =1}^{\mathsf{DOF}}{\sigma _{\ell}}}\sum_{q=0}^{\infty}{\frac{\psi _q\left( \frac{x}{\overline{\gamma }_{\mathrm{b}}} \right) ^{\mathsf{DOF}+q-1}e^{-\frac{x}{\overline{\gamma }_{\mathrm{b}}\sigma _{\min}}}}{\sigma _{\min}^{q}\Gamma (\mathsf{DOF}+q)},}
\end{equation}
where $x>0$. $\Gamma \left( z \right)$ denotes the Gamma function, defined as $\Gamma \left( z \right) =\int_0^{\infty}{e^{-t}t^{z-1}\mathrm{d}t}$, and $\sigma _{\min}=\min \left\{ \sigma _{\ell} \right\} _{\ell=1}^{\infty}=\sigma _{\mathsf{DOF}}$. The coefficients $\psi _{q}$ are given by
\begin{equation}
\label{eq25}
\psi _q=\sum_{k=1}^q{\left[ \sum_{\ell =1}^{\mathsf{DOF}}{\left( 1-\sigma _{\min}/\sigma _{\ell} \right) ^k} \right] \frac{\psi _{q-k}}{q},q}=1,2,...,
\end{equation}
with $\psi _{0}=1$.
\vspace{-0.3cm}
\subsection{PDF of the SNR for a Single Eve}
When there is a single Eve, i.e., $K=1$, the channel between Alice and Eve can be represented as follows:
\begin{equation}
\label{eq26}
h_{\mathrm{e}_1}(\mathbf{r},\mathbf{t})=\frac{\sigma _{h_{\mathrm{e}_1}}^{2}}{(2\pi )^2}\int_{-k_0}^{k_0}{{e}^{\mathrm{j}\kappa _zz}H_{\mathrm{e}_1}(\kappa _z)\mathrm{d}\kappa _z}\triangleq g_{\mathrm{e}_1}(z).
\end{equation}

Using a similar approach as in the analysis of Bob's instantaneous SNR, the channel between Alice and Eve can be decomposed as follows:
\begin{equation}
\label{eq27}
g_{\mathrm{e}_1}(z)\overset{d}{=}\overline{g}_{\mathrm{e}_1}(z)=\sigma _{h_{\mathrm{e}_1}}^{2}\sum_{\ell =1}^{\infty}{\int_{-\frac{L}{2}}^{\frac{L}{2}}{\sigma _{\ell}^{\frac{1}{2}}\phi _{\ell}(z)\phi _{\ell}^{*}(z^{\prime})\overline{W}_{\mathrm{e}_1}(z^{\prime})\mathrm{d}z^{\prime}.}}
\end{equation}

The effective noiseless signal received by Eve can be expressed as follows:
\begin{equation}
\label{eq28}
\begin{aligned}
\overline{y}_{\mathrm{e}_1}&=\sqrt{P}s\cdot \frac{\int_{\mathcal{A}}{h_{\mathrm{e}_1}\left( s \right) h_{\mathrm{b}}^{*}\left( s \right) ds}}{\sqrt{\int_{\mathcal{A}}{\left| h_{\mathrm{b}}\left( s \right) \right|^2ds}}}\\
&\overset{d}{=}\sqrt{P}s\cdot \frac{\sigma _{h_{\mathrm{e}_1}}\sum_{\ell =1}^{\infty}{\sigma _{\ell}^{\frac{1}{2}}\Phi _{{\mathrm{e}_1},\ell}\int_{\mathcal{A}}{\phi _{\ell}(s)h_{\mathrm{b}}^{*}\left( s \right) ds}}}{\sqrt{\int_{\mathcal{A}}{\left| h_{\mathrm{b}}\left( s \right) \right|^2ds}}}.
\end{aligned}
\end{equation}

Each term $\overline{y}_{\mathrm{e}_1,\ell}$, $\ell \in \left\{ 1,2,...,\infty \right\} $, can be expressed as follows:
\begin{equation}
\label{eq29}
\begin{aligned}
\overline{y}_{\mathrm{e}_1,\ell}&\overset{d}{=}\Phi _{\mathrm{e}_1,\ell}\cdot \frac{\sqrt{P}s \sigma _{h_{\mathrm{e}_1}}\sigma _{\ell}^{\frac{1}{2}}\int_{\mathcal{A}}{\phi _{\ell}(s)h_{\mathrm{b}}^{*}\left( s \right) ds}}{\sqrt{\int_{\mathcal{A}}{\left| h_{\mathrm{b}}\left( s \right) \right|^2ds}}}
\\
&\overset{d}{=}\Phi _{\mathrm{e}_1,\ell}\cdot \frac{\sqrt{P}s \sigma _{h_{\mathrm{e}_1}}\sigma _{\ell}^{\frac{1}{2}}\int_{\mathcal{A}}{\phi _{\ell}(s)\sum_{p=1}^{\infty}{\sigma _{p}^{\frac{1}{2}}\phi _{p}^{*}(s)\Phi _{\mathrm{b},p}^{*}ds}}}{\sqrt{\int_{\mathcal{A}}{\left| \sum_{n=1}^{\infty}{\sigma _{n}^{\frac{1}{2}}\phi _n(s)\Phi _{\mathrm{b},n}} \right|^2ds}}}.
\end{aligned}
\end{equation}
A realization of ${h_{\mathrm{b}}}$ is denoted by ${\underline h _{\mathrm{b}}}$, while the corresponding realizations of $\Phi _{\mathrm{b},p}$ and $\Phi _{\mathrm{b},n}$ are denoted by $\underline{\Phi }_{\mathrm{b},p}$ and $\underline{\Phi }_{\mathrm{b},n}$, respectively. It then follows that
\begin{equation}
\label{eq30}
\overline{y}_{\mathrm{e}_1,\ell}|\underline h_{\mathrm{b}}\overset{d}{=}\Phi _{\mathrm{e}_1,\ell}\cdot \frac{\sqrt{P}s \sigma _{h_{\mathrm{e}_1}}\sigma _{\ell}^{\frac{1}{2}}\int_{\mathcal{A}}{\phi _{\ell}(s)\sum_{p=1}^{\infty}{\sigma _{p}^{\frac{1}{2}}\phi _{p}^{*}(s)\underline{\Phi }_{\mathrm{b},p}^{*}ds}}}{\sqrt{\int_{\mathcal{A}}{\left| \sum_{n=1}^{\infty}{\sigma _{n}^{\frac{1}{2}}\phi _n(s)\underline{\Phi }_{\mathrm{b},n}} \right|^2ds}}}.
\end{equation}
It can be observed that $\overline{y}_{\mathrm{e}_1,\ell}$ is a zero-mean complex Gaussian random variable. The variance of $\overline{y}_{\mathrm{e}_1,\ell}$ conditioned on $\underline h_{\mathrm{b}}$ is given by
\begin{equation}
\label{eq31}
\begin{aligned}
\mathrm{var}\left( \overline{y}_{\mathrm{e}_1,\ell}|\underline h_{\mathrm{b}} \right) &=\frac{P \sigma _{h_{\mathrm{e}_1}}^{2}\sigma _{\ell}\left| \int_{\mathcal{A}}{\phi _{\ell}(s)\sum_{p=1}^{\infty}{\sigma _{p}^{\frac{1}{2}}\phi _{p}^{*}(s)\underline{\Phi }_{\mathrm{b},p}^{*}ds}} \right|^2}{\int_{\mathcal{A}}{\left| \sum_{n=1}^{\infty}{\sigma _{n}^{\frac{1}{2}}\phi _n(s)\underline{\Phi }_{\mathrm{b},n}} \right|^2ds}}
\\
&=\frac{P \sigma _{h_{\mathrm{e}_1}}^{2}\sigma _{\ell}^{2}\left| \underline{\Phi }_{\mathrm{b},\ell} \right|^2}{\sum_{n=1}^{\infty}{\sigma _n\left| \underline{\Phi }_{\mathrm{b},n} \right|^2}}.
\end{aligned}
\end{equation}
Since $\Phi _{\mathrm{b},\ell}$, $\ell =1,2,...,\infty$, are independent zero-mean complex Gaussian random variables, the variance of $\overline{y}_{\mathrm{e}_1}$ conditioned on $\underline h_{\mathrm{b}}$ can be expressed as follows:
\begin{equation}
\label{eq32}
\mathrm{var}\left( \overline{y}_{\mathrm{e}_1}|\underline h_{\mathrm{b}} \right) =\frac{P\sigma _{h_{\mathrm{e}_1}}^{2}\sum_{\ell =1}^{\infty}{\sigma _{\ell}^{2}\left| \underline{\Phi }_{\mathrm{b},\ell} \right|^2}}{\sum_{n=1}^{\infty}{\sigma _n\left| \underline{\Phi }_{\mathrm{b},n} \right|^2}}.
\end{equation}
According to Landau's eigenvalue theorem, it follows that
\begin{equation}
\label{eq33}
\mathrm{var}\left( \overline{y}_{\mathrm{e}_1}|\underline h_{\mathrm{b}} \right) \approx \frac{\frac{\lambda}{2}P\sigma _{h_{\mathrm{e}_1}}^{2}\sum_{\ell =1}^{\mathsf{DOF}}{\sigma _{\ell}\left| \underline{\Phi }_{\mathrm{b},\ell} \right|^2}}{\sum_{n=1}^{\mathsf{DOF}}{\sigma _n\left| \underline{\Phi }_{\mathrm{b},n} \right|^2}}=\frac{\lambda}{2}P\sigma _{h_{\mathrm{e}_1}}^{2}.
\end{equation}
It can be observed that the variance of $\overline{y}_{\mathrm{e}_1}$ conditioned on $\underline h_{\mathrm{b}}$, can be regarded as approximately independent of $\underline h_{\mathrm{b}}$. Thus, $\overline{y}_{\mathrm{e}_1}$ is a zero-mean complex Gaussian random variable with variance $\frac{\lambda}{2}P\sigma _{h_{\mathrm{e}_1}}^{2}$ and is approximately independent of $h_{\mathrm{b}}$. Consequently, PDF of Eve's instantaneous SNR, ${\rho _{\mathrm{S,e}}}$, can be approximated as follows:
\begin{equation}
\label{eq34}
f_{\rho _\mathrm{S,e}}\left( x \right) \approx \left\{ \begin{array}{c}
	0,x<0\\
	\frac{1}{\overline{\gamma }_\mathrm{e}}e^{-\frac{1}{\overline{\gamma }_\mathrm{e}}x},x\geqslant 0,\\
\end{array} \right.
\end{equation}
where $\overline{\gamma }_{\mathrm{e}}=\frac{\lambda P\sigma _{h_{\mathrm{e}_1}}^{2}}{2\sigma _{\mathrm{e}_1}^{2}}$ holds.
\vspace{-0.3cm}
\subsection{PDF of the SNR for Multiple Independent Eves}
Under the multiple independent Eves scenario, it is assumed that the eavesdropping channels are i.i.d., while the noise power at each Eve is the same, i.e., $\sigma _{\mathrm{e}}^{2}$. Then, the cumulative distribution function (CDF) of the highest instantaneous SNR among Eves \cite{wang2007secrecy} can be expressed as follows:
\begin{equation}
\label{eq35}
\begin{array}{l}
F_{\rho _\mathrm{M,e}}=\left( 1-e^{-\frac{1}{\overline{\gamma }_\mathrm{e}}x} \right) ^K,x\geqslant 0.
\end{array}
\end{equation}
Then, PDF of the highest instantaneous SNR among Eves is given by
\begin{equation}
\label{eq36}
\begin{array}{l}
f_{\rho _\mathrm{M,e}}=K\left( 1-e^{-\frac{1}{\overline{\gamma }_\mathrm{e}}x} \right) ^{K-1}\frac{1}{\overline{\gamma }_\mathrm{e}}e^{-\frac{1}{\overline{\gamma }_\mathrm{e}}x}.
\end{array}
\end{equation}

\subsection{PDF of the SNR for Multiple Collaborative Eves}
Under the multiple collaborative Eves scenario, all Eves form a distribute antenna system. Maximum-ratio combining (MRC) is employed at Eves' side to maximize the received signal power. It is also assumed that the eavesdropping channels are i.i.d., and that the noise power at each Eve is the same. After applying MRC to the received signal in (\ref{eq5}), the processed signal at Eve $k$ can be expressed as follows:
\begin{equation}
\label{eq37}
\begin{small}
\begin{array}{l}
y_{\mathrm{e}_k}\left( \mathbf{r} \right) =\sqrt{P}s\left( \frac{\int_{\mathcal{A}}{h_{\mathrm{e}_k}\left( \mathbf{r},\mathbf{t} \right) h_{\mathrm{b}}^{*}\left( \mathbf{r},\mathbf{t} \right) d\mathbf{t}}}{\sqrt{\int_{\mathcal{A}}{\left| h_{\mathrm{b}}\left( \mathbf{r},\mathbf{t} \right) \right|^2d\mathbf{t}}}} \right) ^*\frac{\int_{\mathcal{A}}{h_{\mathrm{e}_k}\left( \mathbf{r},\mathbf{t} \right) h_{\mathrm{b}}^{*}\left( \mathbf{r},\mathbf{t} \right) d\mathbf{t}}}{\sqrt{\int_{\mathcal{A}}{\left| h_{\mathrm{b}}\left( \mathbf{r},\mathbf{t} \right) \right|^2d\mathbf{t}}}}
\\
+\left( \frac{\int_{\mathcal{A}}{h_{\mathrm{e}_k}\left( \mathbf{r},\mathbf{t} \right) h_{\mathrm{b}}^{*}\left( \mathbf{r},\mathbf{t} \right) d\mathbf{t}}}{\sqrt{\int_{\mathcal{A}}{\left| h_{\mathrm{b}}\left( \mathbf{r},\mathbf{t} \right) \right|^2d\mathbf{t}}}} \right) ^*n_{\mathrm{e}_k}\left( \mathbf{r} \right)
\\
=\sqrt{P}s\left| \frac{\int_{\mathcal{A}}{h_{\mathrm{e}_k}\left( \mathbf{r},\mathbf{t} \right) h_{\mathrm{b}}^{*}\left( \mathbf{r},\mathbf{t} \right) d\mathbf{t}}}{\sqrt{\int_{\mathcal{A}}{\left| h_{\mathrm{b}}\left( \mathbf{r},\mathbf{t} \right) \right|^2d\mathbf{t}}}} \right|^2+\left( \frac{\int_{\mathcal{A}}{h_{\mathrm{e}_k}\left( \mathbf{r},\mathbf{t} \right) h_{\mathrm{b}}^{*}\left( \mathbf{r},\mathbf{t} \right) d\mathbf{t}}}{\sqrt{\int_{\mathcal{A}}{\left| h_{\mathrm{b}}\left( \mathbf{r},\mathbf{t} \right) \right|^2d\mathbf{t}}}} \right) ^*n_{\mathrm{e}_k}\left( \mathbf{r} \right) .
\end{array}
\end{small}
\end{equation}
Therefore, the instantaneous SNR, $\rho _\mathrm{MC,e}$, is the sum of i.i.d. exponential random variables. Consequently, $\rho _\mathrm{MC,e}$ follows a Gamma distribution, and its PDF can be expressed as follows:
\begin{equation}
\label{eq38}
f_{\rho _{\mathrm{MC},\mathrm{e}}}\left( x \right) \approx \left\{ \begin{array}{c}
	0,x<0\\
	\frac{x^{K-1}e^{-\frac{x}{\overline{\gamma }_{\mathrm{e}}}}}{\overline{\gamma }_{\mathrm{e}}^{K}\Gamma \left( K \right)},x\geqslant 0.\\
\end{array} \right.
\end{equation}
\vspace{-0.3cm}
\section{Secrecy Rate and SOP of CAPA Systems}
\subsection{Secrecy Rate and SOP Under a Single Eve}
The mutual information between Alice and Bob, and between Alice and Eve, can be expressed as follows:
\begin{equation}
\label{eq39}
I\left( x;y_{\mathrm{b}} \right) =\log _2\left( 1+\rho _{\mathrm{b}} \right) ,
\\
I\left( x;y_{\mathrm{S},\mathrm{e}} \right) =\log _2\left( 1+\rho _{\mathrm{S},\mathrm{e}} \right) .
\end{equation}

The secrecy rate of the system is given by
\begin{equation}
\label{eq40}
R_{\mathrm{S}}=\mathbb{E} \left\{ \max \left( \log _2\left( 1+\rho _{\mathrm{b}} \right) -\log _2\left( 1+\rho _{\mathrm{S},\mathrm{e}} \right) ,0 \right) \right\} .
\end{equation}
Since $h_{\mathrm{b}}(\mathbf{r},\mathbf{t})$ and $h_{\mathrm{e}}(\mathbf{r},\mathbf{t})$ are approximately independent, as shown in (\ref{eq33}), the instantaneous SNRs $\rho _{\mathrm{b}}$ and $\rho _{\mathrm{e}}$ can likewise be considered approximately independent. Accordingly, the secrecy rate of the CAPA system under a single Eve can be expressed as follows:
\begin{equation}
\label{eq41}
\begin{small}
\begin{array}{l}
R_{\mathrm{S}}=\int_0^{\infty}{\int_{\rho _{\mathrm{S},\mathrm{e}}}^{\infty}{\left[ \log _2\left( 1+\rho _{\mathrm{b}} \right) -\log _2\left( 1+\rho _{\mathrm{S},\mathrm{e}} \right) \right] f_{\rho _{\mathrm{b}}}f_{\rho _{\mathrm{S},\mathrm{e}}}}d\rho _{\mathrm{b}}d\rho _{\mathrm{S},\mathrm{e}}.}
\end{array}
\end{small}
\end{equation}
After straightforward mathematical derivations, the final result of $R_{\mathrm{S}}$ is provided in Appendix B.

To evaluate the high-SNR slope and high-SNR power offset \cite{shamai2002impact}, the following lemma is introduced.

\textbf{Lemma 1.} Let $y=\ln \left( x \right) +\mathrm{E}_1\left( x \right) $. As $x\rightarrow 0^+$, we have $y=-\gamma +o\left( x \right) $, where $-\gamma $ denotes the Euler-Mascheroni constant.

\textit{Proof}: As $x\rightarrow 0^+$, the series expansion of $\mathrm{E}_1\left( x \right)$ can be written as
\begin{equation}
\label{eq42}
\mathrm{E}_1\left( x \right) =-\gamma -\ln x+\sum_{k=1}^{\infty}{\frac{\left( -1 \right) ^{k-1}x^k}{k\cdot k!}}.
\end{equation}
Based on $y=\ln \left( x \right) +\mathrm{E}_1\left( x \right) $, we have $y=-\gamma +o\left( x \right) $.
$\hfill\blacksquare$

In Bob's high-SNR region, based on (\ref{eq83}), the secrecy rate can be approximated as follows:
\begin{equation}
\label{eq43}
\begin{array}{l}
\underset{\overline{\gamma }_{\mathrm{b}}\rightarrow \infty}{\lim}R_{\mathrm{S}}\approx \frac{\sigma _{\min}^{\mathsf{DOF}}}{\prod_{\ell =1}^{\mathsf{DOF}}{\sigma _{\ell}}}\sum_{q=0}^{\infty}{\psi _q\left\{ \log _2\left( \overline{\gamma }_{\mathrm{b}}\sigma _{\min} \right) -\frac{1}{\ln 2} \right. \}}
\\
\cdot \left. \left[ e^{\frac{1}{\overline{\gamma }_{\mathrm{e}}}}\mathrm{E}_1\left( \frac{1}{\overline{\gamma }_{\mathrm{e}}} \right) +\gamma -\sum_{p=0}^{\mathsf{DOF}+q-2}{\frac{1}{\left( \mathsf{DOF}+q-1-p \right)}} \right] \right. .
\end{array}
\end{equation}
The high-SNR slope can then be expressed as follows:
\begin{equation}
\label{eq44}
S_{\mathrm{S},\infty}=\underset{\overline{\gamma }_{\mathrm{b}}\rightarrow \infty}{\lim}\frac{R_{\mathrm{S}}}{\log _2\overline{\gamma }_{\mathrm{b}}}\approx \frac{\sigma _{\min}^{\mathsf{DOF}}}{\prod_{\ell =1}^{\mathsf{DOF}}{\sigma _{\ell}}}\sum_{q=0}^{\infty}{\psi _q}.
\end{equation}
Accordingly, the high-SNR power offset is given by
\begin{equation}
\label{eq45}
\begin{array}{l}
\mathcal{L} _{\mathrm{S},\infty}=\underset{\overline{\gamma }_{\mathrm{b}}\rightarrow \infty}{\lim}\left( \log _2\overline{\gamma }_{\mathrm{b}}-\frac{R_{\mathrm{S}}}{S_{\mathrm{S},\infty}} \right)
\\
\approx -\log _2\left( \sigma _{\min} \right) +\frac{1}{\ln 2}\left[ \gamma +e^{\frac{1}{\overline{\gamma }_{\mathrm{e}}}}\mathrm{E}_1\left( \frac{1}{\overline{\gamma }_{\mathrm{e}}} \right) \right.
\\
-\left. \left( \sum_{q=0}^{\infty}{\psi _q\sum_{p=0}^{\mathsf{DOF}+q-2}{\frac{1}{\left( \mathsf{DOF}+q-1-p \right)}}} \right) /\sum_{q=0}^{\infty}{\psi _q} \right] .
\end{array}
\end{equation}

SOP corresponding to the target secrecy rate $R_0$ can be expressed as follows:
\begin{equation}
\label{eq46}
\begin{array}{l}
P_{\mathrm{SOP}_{\mathrm{S}}}\left( R_0 \right) =P\left( R_{\mathrm{S}}<R_0 \right) =P\left( \rho _{\mathrm{b}}<2^{R_0}\left( 1+\rho _{\mathrm{e}} \right) -1 \right)
\\
\approx \frac{\sigma _{\min}^{\mathsf{DOF}}}{\prod_{\ell =1}^{\mathsf{DOF}}{\sigma _{\ell}}}\sum_{q=0}^{\infty}{\psi _q}\left[ 1-e^{-\frac{2^{R_0}-1}{\overline{\gamma }_{\mathrm{b}}\sigma _{\min}}}\sum_{k=0}^{\mathsf{DOF}+q-1}{\frac{1}{\left( \overline{\gamma }_{\mathrm{b}}\sigma _{\min} \right) ^k}} \right.
\\
\cdot\left. \sum_{m=0}^k{\frac{\left( 2^{R_0}-1 \right) ^{k-m}\left( 2^{R_0}\overline{\gamma }_{\mathrm{e}} \right) ^m}{\left( k-m \right) !}\left( \frac{\overline{\gamma }_{\mathrm{b}}\sigma _{\min}}{\overline{\gamma }_{\mathrm{b}}\sigma _{\min}+2^{R_0}\overline{\gamma }_{\mathrm{e}}} \right) ^{m+1}} \right].
\end{array}
\end{equation}

It can then be observed that the CAPA system achieves a diversity order of $\mathsf{DOF}$ for SOP, whereas the array gain is
\begin{equation}
\label{eq47}
\begin{array}{l}
\mathrm{Ag}_{\mathrm{S}}\approx \frac{1}{2^{R_0}\overline{\gamma }_{\mathrm{e}}}\left[ \left( \prod_{\ell =1}^{\mathsf{DOF}}{\sigma _{\ell}} \right) /\left( \psi _0\sum_{m=0}^{\mathsf{DOF}}{\frac{1}{m!}\left( \frac{2^{R_0}-1}{2^{R_0}\overline{\gamma }_{\mathrm{e}}} \right) ^m} \right) \right] ^{\frac{1}{\mathsf{DOF}}}.
\end{array}
\end{equation}
The detailed mathematical derivation is provided in Appendix C.
\vspace{-0.3cm}
\subsection{Secrecy Rate and SOP Under Multiple Independent Eves}
The mutual information between Alice and the Eve with the highest SNR is
\begin{equation}
\label{eq48}
\begin{array}{l}
I\left( x;y_{\mathrm{M},\mathrm{e}} \right) =\log _2\left( 1+\rho _{\mathrm{M},\mathrm{e}} \right) .
\end{array}
\end{equation}
The secrecy rate of the system in the presence of multiple independent Eves can be expressed as follows:
\begin{equation}
\label{eq49}
R_{\mathrm{M}}=\mathbb{E} \left\{ \max \left( \log _2\left( 1+\rho _{\mathrm{b}} \right) -\log _2\left( 1+\rho _{\mathrm{M},\mathrm{e}} \right) ,0 \right) \right\} .
\end{equation}
Since $h_{\mathrm{b}}(\mathbf{r},\mathbf{t})$ is independent of $h_{\mathrm{e}_k}(\mathbf{r},\mathbf{t})$, $\rho _\mathrm{b}$ and $\rho _{\mathrm{M},\mathrm{e}}$ are also independent. Consequently, the secrecy rate can be expressed as follows:
\begin{equation}
\label{eq50}
\begin{small}
\begin{array}{l}
R_{\mathrm{M}}=\int_0^{\infty}{\int_{\rho _{\mathrm{M},\mathrm{e}}}^{\infty}{\left[ \log _2\left( 1+\rho _{\mathrm{b}} \right) -\log _2\left( 1+\rho _{\mathrm{M},\mathrm{e}} \right) \right] f_{\rho _{\mathrm{b}}}f_{\rho _{\mathrm{M},\mathrm{e}}}d\rho _{\mathrm{b}}d\rho _{\mathrm{M},\mathrm{e}},}}
\end{array}
\end{small}
\end{equation}
After some straightforward mathematical derivations, the final result of (\ref{eq50}) is presented in Appendix D.

To evaluate the high-SNR slope and the high-SNR power offset in the multiple independent Eves scenario, we introduce the following lemma.

\textbf{Lemma 2.} Let $K\geqslant 1$, $K\subseteq \mathbb{Z} ^+$, it holds that $K\sum_{d=0}^{K-1}{\left( \begin{array}{c}
	K-1\\
	d\\
\end{array} \right) \left( -1 \right) ^d}\frac{1}{d+1}=1$.

\textit{Proof}: The left-hand side can be expressed as follows:
\begin{equation}
\label{eq51}
\begin{array}{l}
K\sum_{d=0}^{K-1}{\left( \begin{array}{c}
	K-1\\
	d\\
\end{array} \right) \left( -1 \right) ^d\int_0^1{x^ddx}}
\\
=K\int_0^1{\sum_{d=0}^{K-1}{\left( \begin{array}{c}
	K-1\\
	d\\
\end{array} \right) \left( -x \right) ^ddx}}
\\
=K\int_0^1{\left( 1-x \right) ^{K-1}dx}
\\
=1.
\end{array}
\end{equation}
This completes the proof.
$\hfill\blacksquare$

Based on the result in (\ref{eq96}), in Bob's high-SNR region, the secrecy rate can be expressed as follows:
\begin{equation}
\label{eq52}
\begin{array}{l}
\underset{\overline{\gamma }_{\mathrm{b}}\rightarrow \infty}{\lim}R_{\mathrm{M}}\approx \frac{\sigma _{\min}^{\mathsf{DOF}}}{\prod_{\ell =1}^{\mathsf{DOF}}{\sigma _{\ell}}}\sum_{q=0}^{\infty}{\psi _q\left\{ \log _2\left( \overline{\gamma }_{\mathrm{b}}\sigma _{\min} \right) \right.}
\\
-\frac{1}{\ln 2}\left( \gamma -\sum_{p=0}^{\mathsf{DOF}+q-2}{\frac{1}{\left( \mathsf{DOF}+q-1-p \right)}} \right)
\\
\left. -\frac{K}{\ln 2}\sum_{a=0}^{K-1}{\left( \begin{array}{c}
	K-1\\
	a\\
\end{array} \right) \frac{\left( -1 \right) ^a}{1+a}e^{\frac{1+a}{\overline{\gamma }_{\mathrm{e}}}}\mathrm{E}_1\left( \frac{1+a}{\overline{\gamma }_{\mathrm{e}}} \right)} \right\} .
\end{array}
\end{equation}
Accordingly, the high-SNR slope in the presence of multiple independent Eves is
\begin{equation}
\label{eq53}
S_{\mathrm{M},\infty}=\underset{\overline{\gamma }_{\mathrm{b}}\rightarrow \infty}{\lim}\frac{R_{\mathrm{M}}}{\log _2\overline{\gamma }_{\mathrm{b}}}\approx \frac{\sigma _{\min}^{\mathsf{DOF}}}{\prod_{\ell =1}^{\mathsf{DOF}}{\sigma _{\ell}}}\sum_{q=0}^{\infty}{\psi _q}.
\end{equation}
As shown in (\ref{eq44}) and (\ref{eq53}), the high-SNR slopes under a single Eve and multiple independent Eves are the same. Meanwhile, the high-SNR power offset is given by
\begin{equation}
\label{eq54}
\begin{small}
\begin{array}{l}
\mathcal{L} _{\mathrm{M},\infty}=\underset{\overline{\gamma }_{\mathrm{b}}\rightarrow \infty}{\lim}\left( \log _2\overline{\gamma }_{\mathrm{b}}-\frac{R_{\mathrm{M}}}{S_{\mathrm{M},\infty}} \right)
\\
\approx -\log _2\left( \sigma _{\min} \right) +\frac{1}{\ln 2}\left[ \gamma +K\sum_{a=0}^{K-1}{\left( \begin{array}{c}
	K-1\\
	a\\
\end{array} \right) \frac{\left( -1 \right) ^a}{1+a}e^{\frac{1+a}{\overline{\gamma }_{\mathrm{e}}}}} \right.
\\
\cdot \left. \mathrm{E}_1\left( \frac{1+a}{\overline{\gamma }_{\mathrm{e}}} \right) -\left( \sum_{q=0}^{\infty}{\psi _q\sum_{p=0}^{\mathsf{DOF}+q-2}{\frac{1}{\left( \mathsf{DOF}+q-1-p \right)}}} \right) /\sum_{q=0}^{\infty}{\psi _q} \right] .
\end{array}
\end{small}
\end{equation}

To compare the high-SNR power offset for the cases of a single Eve and multiple independent Eves, the following lemma is introduced.

\textbf{Lemma 3.} Let $K\geqslant 1$ with $K\subseteq \mathbb{Z} ^+$, and define $y\left( K \right) =K\sum_{a=0}^{K-1}{\left( \begin{array}{c}
	K-1\\
	a\\
\end{array} \right) \frac{\left( -1 \right) ^a}{1+a}e^{\frac{1+a}{\overline{\gamma }_{\mathrm{e}}}}\mathrm{E}_1\left( \frac{1+a}{\overline{\gamma }_{\mathrm{e}}} \right)}$. Then, $y\left( K \right)$ is an increasing function of $K$.

\textit{Proof}: $y\left( K \right)$ can be simplified as follows:
\begin{equation}
\label{eq55}
\begin{small}
\begin{array}{l}
y\left( K \right) =\sum_{b=1}^K{\left( -1 \right) ^{b-1}\left( \begin{array}{c}
	K\\
	b\\
\end{array} \right) e^{\frac{b}{\overline{\gamma }_e}}\mathrm{E}_1\left( \frac{b}{\overline{\gamma }_e} \right) .}
\end{array}
\end{small}
\end{equation}
For $K\geqslant 1$ with $K\subseteq \mathbb{Z} ^+$, we have
\begin{equation}
\label{eq56}
\begin{small}
\begin{array}{l}
y\left( K+1 \right) -y\left( K \right) =\sum_{c=0}^K{\left( -1 \right) ^c\left( \begin{array}{c}
	K\\
	c\\
\end{array} \right) \int_0^{\infty}{\frac{e^{-\left( 1+c \right) y/\overline{\gamma }_e}}{1+y}dy}}
\\
=\int_0^{\infty}{\frac{e^{-y/\overline{\gamma }_e}}{1+y}\left( 1-e^{-y/\overline{\gamma }_e} \right) ^Kdy}>0.
\end{array}
\end{small}
\end{equation}
Thus, $y\left( K \right)$ is an increasing function of $K$ for $K\geqslant 1$ with $K\subseteq \mathbb{Z} ^+$. This completes the proof. $\hfill\blacksquare$

Since $\mathcal{L} _{\mathrm{S},\infty}$ is a special case of $\mathcal{L} _{\mathrm{M},\infty}$ with $K=1$, Lemma 3 together with (\ref{eq45}) and (\ref{eq54}) implies that, the CAPA system with multiple independent Eves exhibits a larger high-SNR power offset than that with a single Eve for $K\geqslant 2$.

Under the multiple independent Eves scenario, SOP under the target secrecy rate $R_0$ can be expressed as follows:
\begin{equation}
\label{eq57}
\begin{array}{l}
P_{\mathrm{SOP}_{\mathrm{M}}}\left( R_0 \right) =P\left( R_{\mathrm{S}}<R_0 \right) =P\left( \rho _{\mathrm{b}}<2^{R_0}\left( 1+\rho _{\mathrm{M},\mathrm{e}} \right) -1 \right)
\\
\approx \frac{K\sigma _{\min}^{\mathsf{DOF}}}{\prod_{\ell =1}^{\mathsf{DOF}}{\sigma _{\ell}}}\sum_{n=0}^{K-1}{\left( \begin{array}{c}
	K-1\\
	n\\
\end{array} \right) \left( -1 \right) ^n}\sum_{q=0}^{\infty}{\psi _q}\left[ \frac{1}{n+1}- \right.
\\
e^{-\frac{2^{R_0}-1}{\overline{\gamma }_{\mathrm{b}}\sigma _{\min}}}\sum_{k=0}^{\mathsf{DOF}+q-1}{\frac{1}{\left( \overline{\gamma }_{\mathrm{b}}\sigma _{\min} \right) ^k}\sum_{m=0}^k{\frac{\left( 2^{R_0}-1 \right) ^{k-m}\left( 2^{R_0}\overline{\gamma }_{\mathrm{e}} \right) ^m}{\left( k-m \right) !}}}
\\
\cdot \left. \left( \frac{\overline{\gamma }_{\mathrm{b}}\sigma _{\min}}{\left( n+1 \right) \overline{\gamma }_{\mathrm{b}}\sigma _{\min}+2^{R_0}\overline{\gamma }_{\mathrm{e}}} \right) ^{m+1} \right].
\end{array}
\end{equation}

Then, it can be obtained that the diversity order of the CAPA system in the presence of multiple independent Eves is $\mathsf{DOF}$, while the array gain of the system is
\begin{equation}
\label{eq58}
\begin{aligned}
\mathrm{Ag}_{\mathrm{M}}\approx \frac{1}{2^{R_0}\overline{\gamma }_{\mathrm{e}}}\left[ \left( \prod_{\ell =1}^{\mathsf{DOF}}{\sigma _{\ell}} \right) / \right. \left( K\psi _0\sum_{m=0}^{\mathsf{DOF}}{\frac{1}{m!}\left( \frac{2^{R_0}-1}{2^{R_0}\overline{\gamma }_{\mathrm{e}}} \right) ^m} \right.
\\
\cdot \left. \left. \sum_{n=0}^{K-1}{\left( \begin{array}{c}
	K-1\\
	n\\
\end{array} \right) \left( -1 \right) ^n\left( \frac{1}{n+1} \right) ^{\mathsf{DOF}-m+1}} \right) \right] ^{\frac{1}{\mathsf{DOF}}}.
\end{aligned}
\end{equation}
The details of the derivation are provided in Appendix E.

The following lemma is introduced to facilitate the comparison of the array gain between the single-Eve and multiple independent Eves scenarios.

\textbf{Lemma 4.} Let $K\geqslant 1$ with $K\subseteq \mathbb{Z} ^+$, and define $y\left( K \right) =K\sum_{n=0}^{K-1}{\left( \begin{array}{c}
	K-1\\
	n\\
\end{array} \right) \left( -1 \right) ^n\left( \frac{1}{n+1} \right) ^{\mathsf{DOF}-m+1}}$, where $0\leqslant m\leqslant \mathsf{DOF}$ with $m\subseteq \mathbb{Z}$. Then, $y\left( K \right)=1$ when $m=\mathsf{DOF}$, and $y\left( K \right)$ is an increase function of $K$ when $m<\mathsf{DOF}$.

\textit{Proof}: When $m=\mathsf{DOF}$, we have
\begin{equation}
\label{eq59}
\begin{array}{l}
y\left( K \right) =K\sum_{n=0}^{K-1}{\left( \begin{array}{c}
	K-1\\
	n\\
\end{array} \right) \left( -1 \right) ^n\frac{1}{n+1}}
\\
=\sum_{b=1}^K{\left( \begin{array}{c}
	K\\
	b\\
\end{array} \right) \left( -1 \right) ^{b-1}}=1.
\end{array}
\end{equation}
When $m<\mathsf{DOF}$, let $\alpha =\mathsf{DOF}-m+1$. Then, it follows that
\begin{equation}
\label{eq60}
\begin{array}{l}
y\left( K+1 \right) -y\left( K \right) =\sum_{n=0}^K{\left( -1 \right) ^n\left( \begin{array}{c}
	K\\
	n\\
\end{array} \right)}\left( n+1 \right) ^{1-\alpha}
\\
=\frac{1}{\Gamma \left( \alpha -1 \right)}\int_0^{\infty}{t^{\alpha -2}e^{-t}\sum_{n=0}^K{\left( \begin{array}{c}
	K\\
	n\\
\end{array} \right)}\left( -e^{-t} \right) ^ndt}
\\
=\frac{1}{\Gamma \left( \alpha -1 \right)}\int_0^{\infty}{t^{\alpha -2}e^{-t}\left( 1-e^{-t} \right) ^Kdt}>0.
\end{array}
\end{equation}
Therefore, $y\left( K \right)$ is an increase function of $K$ when $m<\mathsf{DOF}$. This completes the proof. $\hfill\blacksquare$

Since $\mathrm{Ag}_{\mathrm{S}}$ is a special case of $\mathrm{Ag}_{\mathrm{M}}$ when $K=1$, Lemma 4 together with (\ref{eq47}) and (\ref{eq58}) shows that, the CAPA system with multiple independent Eves has a lower array gain than that with a single Eve for $K\geqslant 2$.

\vspace{-0.3cm}
\subsection{Secrecy Rate and SOP Under Multiple Collaborative Eves}
The mutual information between Alice and multiple collaborative Eves is
\begin{equation}
\label{eq61}
\begin{array}{l}
I\left( x;y_{\mathrm{MC},\mathrm{e}} \right) =\log _2\left( 1+\rho _{\mathrm{MC},\mathrm{e}} \right) .
\end{array}
\end{equation}

The secrecy rate of the system in the presence of multiple collaborative Eves can be expressed as follows:
\begin{equation}
\label{eq62}
R_{\mathrm{MC}}=\mathbb{E} \left\{ \max \left( \log _2\left( 1+\rho _{\mathrm{b}} \right) -\log _2\left( 1+\rho _{\mathrm{MC},\mathrm{e}} \right) ,0 \right) \right\} .
\end{equation}

Since $h_{\mathrm{b}}(\mathbf{r},\mathbf{t})$ and $h_{\mathrm{e}_k}(\mathbf{r},\mathbf{t})$ are independent, $\rho _{\mathrm{b}}$ and $\rho _{\mathrm{MC},\mathrm{e}}$ are also independent. Accordingly, the secrecy rate of the CAPA system in the presence of multiple collaborative Eves can be expressed as follows:
\begin{equation}
\label{eq63}
\begin{aligned}
R_{\mathrm{MC}}=&\int_0^{\infty}{\int_{\rho _{\mathrm{MC},\mathrm{e}}}^{\infty}{\left[ \log _2\left( 1+\rho _{\mathrm{b}} \right) -\log _2\left( 1+\rho _{\mathrm{MC},\mathrm{e}} \right) \right]}}
\\
&\cdot f_{\rho _{\mathrm{b}}}f_{\rho _{\mathrm{MC},\mathrm{e}}}d\rho _{\mathrm{b}}d\rho _{\mathrm{MC},\mathrm{e}}.
\end{aligned}
\end{equation}

The result of (\ref{eq63}) after mathematical manipulations is provided in Appendix F.

In Bob's high-SNR region, we have
\begin{equation}
\label{eq64}
\begin{array}{l}
\underset{\overline{\gamma }_{\mathrm{b}}\rightarrow \infty}{\lim}R_{\mathrm{MC}}\approx \frac{\sigma _{\min}^{\mathsf{DOF}}}{\prod_{\ell =1}^{\mathsf{DOF}}{\sigma _{\ell}}}\sum_{q=0}^{\infty}{\psi _q\left\{ \log _2\left( \overline{\gamma }_{\mathrm{b}}\sigma _{\min} \right) -\frac{1}{\ln 2} \right.}
\\
\cdot \left. \left( e^{\frac{1}{K\overline{\gamma }_{\mathrm{e}}}}\mathrm{E}_1\left( \frac{1}{K\overline{\gamma }_{\mathrm{e}}} \right) +\gamma -\sum_{p=0}^{\mathsf{DOF}+q-2}{\frac{1}{\left( \mathsf{DOF}+q-1-p \right)}} \right) \right\} .
\end{array}
\end{equation}
Accordingly, the high-SNR slope in the presence of multiple collaborative Eves is
\begin{equation}
\label{eq65}
S_{\mathrm{MC},\infty}=\underset{\overline{\gamma }_{\mathrm{b}}\rightarrow \infty}{\lim}\frac{R_{\mathrm{MC}}}{\log _2\bar{\gamma}_{\mathrm{b}}}\approx \frac{\sigma _{\min}^{\mathsf{DOF}}}{\prod_{\ell =1}^{\mathsf{DOF}}{\sigma _{\ell}}}\sum_{q=0}^{\infty}{\psi _q.}
\end{equation}
It can be found from (\ref{eq44}), (\ref{eq53}), and (\ref{eq65}) that the secrecy rates under three different scenarios have the same high-SNR slope. Meanwhile, the corresponding high-SNR power offset is
\begin{equation}
\label{eq66}
\begin{array}{l}
\mathcal{L} _{\mathrm{MC},\infty}=\underset{\overline{\gamma }_{\mathrm{b}}\rightarrow \infty}{\lim}\left( \log _2\overline{\gamma }_{\mathrm{b}}-\frac{R_{\mathrm{MC}}}{S_{\mathrm{MC},\infty}} \right)
\\
\approx -\log _2\left( \sigma _{\min} \right) +\frac{1}{\ln 2}\left[ \gamma +e^{\frac{1}{K\overline{\gamma }_{\mathrm{e}}}}\mathrm{E}_1\left( \frac{1}{K\overline{\gamma }_{\mathrm{e}}} \right) \right.
\\
-\left. \left( \sum_{q=0}^{\infty}{\psi _q\sum_{p=0}^{\mathsf{DOF}+q-2}{\frac{1}{\left( \mathsf{DOF}+q-1-p \right)}}} \right) /\sum_{q=0}^{\infty}{\psi _q} \right] .
\end{array}
\end{equation}

To compare the high-SNR power offset for the cases of multiple independent Eves and multiple collaborative Eves, the following lemma is introduced.

\textbf{Lemma 5.} Let $K\geqslant 2$ with $K\subseteq \mathbb{Z} ^+$, and define $y\left( K \right) =e^{\frac{1}{K\overline{\gamma }_{\mathrm{e}}}}\mathrm{E}_1\left( \frac{1}{K\overline{\gamma }_{\mathrm{e}}} \right) -K\sum_{a=0}^{K-1}{\left( \begin{array}{c}
	K-1\\
	a\\
\end{array} \right) \frac{\left( -1 \right) ^a}{1+a}e^{\frac{1+a}{\overline{\gamma }_{\mathrm{e}}}}\mathrm{E}_1\left( \frac{1+a}{\overline{\gamma }_{\mathrm{e}}} \right)}$. Then, $y\left( K \right)>0$ holds.

\textit{Proof}: Let $f\left( x \right) =e^x\mathrm{E}_1\left( x \right) $ for $x>0$. We have
\begin{equation}
\label{eq67}
\begin{array}{l}
f^{\prime}\left( x \right) =e^x\int_x^{\infty}{\frac{e^{-u}}{u}du}-\frac{1}{x}<e^x\frac{1}{x}\int_x^{\infty}{e^{-u}du}-\frac{1}{x}=0.
\end{array}
\end{equation}
Thus, $f\left( x \right)$ is a decreasing function on $x>0$. Then, we have
\begin{equation}
\label{eq68}
\begin{array}{l}
y\left( K \right) =f\left( \frac{1}{K\overline{\gamma }_e} \right) -K\sum_{a=0}^{K-1}{\left( \begin{array}{c}
	K-1\\
	a\\
\end{array} \right) \frac{\left( -1 \right) ^a}{1+a}f\left( \frac{1+\alpha}{\overline{\gamma }_e} \right)}
\\
<f\left( \frac{1}{K\overline{\gamma }_e} \right) -K\sum_{a=0}^{K-1}{\left( \begin{array}{c}
	K-1\\
	a\\
\end{array} \right) \frac{\left( -1 \right) ^a}{1+a}}f\left( \frac{1}{K\overline{\gamma }_e} \right) =0.
\end{array}
\end{equation}
This completes the proof. $\hfill\blacksquare$

Based on Lemma 5 and expressions of $\mathcal{L} _{\mathrm{M},\infty}$ and $\mathcal{L} _{\mathrm{MC},\infty}$ in (\ref{eq54}) and (\ref{eq66}), it follows that the CAPA system with multiple collaborative Eves exhibits a larger high-SNR power offset than that with multiple independent Eves for $K\geqslant 2$.

For the target secrecy rate $R_0$, SOP in the presence of multiple collaborative Eves can be expressed as follows:
\begin{equation}
\label{eq69}
\begin{array}{l}
$$P_{\mathrm{SOP}_{\mathrm{MC}}}\left( R_0 \right) =P\left( R_{\mathrm{MC}}<R_0 \right)
\\
=P\left( \rho _{\mathrm{b}}<2^{R_0}\left( 1+\rho _{\mathrm{e},\mathrm{MC}} \right) -1 \right)
\\
\approx \frac{\sigma _{\min}^{\mathsf{DOF}}}{\prod_{\ell =1}^{\mathsf{DOF}}{\sigma _{\ell}}}\sum_{q=0}^{\infty}{\psi _q\left( 1-\frac{1}{\left( K-1 \right) !}e^{-\frac{2^{R_0}-1}{\overline{\gamma }_{\mathrm{b}}\sigma _{\min}}} \right.}
\\
\cdot \sum_{k=0}^{\mathsf{DOF}+q-1}{\frac{1}{k!\left( \overline{\gamma }_{\mathrm{b}}\sigma _{\min} \right) ^k}\sum_{m=0}^k{\left( \begin{array}{c}
	k\\
	m\\
\end{array} \right) \left( 2^{R_0}-1 \right) ^{k-m}}}
\\
\cdot \left. \left( 2^{R_0}\overline{\gamma }_{\mathrm{e}} \right) ^m\left( \frac{\overline{\gamma }_{\mathrm{b}}\sigma _{\min}}{\overline{\gamma }_{\mathrm{b}}\sigma _{\min}+2^{R_0}\overline{\gamma }_{\mathrm{e}}} \right) ^{K+m}\left( K+m-1 \right) ! \right) $$.
\end{array}
\end{equation}

Then, it follows that the diversity order of the CAPA system in the presence of multiple collaborative Eves is $\mathsf{DOF}$, whereas the array gain of the system is
\begin{equation}
\label{eq70}
\begin{aligned}
\mathrm{Ag}_{\mathrm{MC}}\approx& \frac{1}{2^{R_0}\overline{\gamma }_{\mathrm{e}}}\left[ \left( \prod_{\ell =1}^{\mathsf{DOF}}{\sigma _{\ell}} \right) /\left( \psi _0\sum_{m=0}^{\mathsf{DOF}}{\frac{1}{m!}} \right. \right.
\\
&\cdot\left. \left. \left( \frac{2^{R_0}-1}{2^{R_0}\overline{\gamma }_{\mathrm{e}}} \right) ^m\left( \begin{array}{c}
	\mathsf{DOF}-m+K-1\\
	K-1\\
\end{array} \right) \right) \right] ^{\frac{1}{\mathsf{DOF}}}.
\end{aligned}
\end{equation}
The details are provided in Appendix G. Therefore, it follows that the diversity orders of the CAPA system in all three scenarios are equal to $\mathsf{DOF}$.
\begin{table}[h!t]
\scriptsize
   \centering%\multicolumn{3}{ r}
   \caption{Simulation parameters}
    \begin{center}
     \begin{tabular}{||c c||}
     \hline
     Parameters & Values \\
     \hline
     SNR of Bob & 20 dB\\
     \hline
     SNR of Eve & 20 dB \\
     \hline
     $\lambda $ & 0.1249 m \\
     \hline
     $L$ & $40\lambda$ \\
     \hline
     $Q$ & 160 \\
     \hline
     $T$ & 1000 \\
     \hline
     Number of Eves & 5 \\
     \hline
     $R_0$ & 3 \\
     \hline
    \end{tabular}
    \end{center}
    \vspace{-0.5cm}
\label{tab1}
\end{table}

The following lemma is introduced to facilitate the comparison of the array gain between multiple independent Eves and multiple collaborative Eves scenarios.

\textbf{Lemma 6.} Let $K\geqslant 2$ with $K\subseteq \mathbb{Z} ^+$, and define $y\left( K \right) =\left( \begin{array}{c}
	\mathsf{DOF}-m+K-1\\
	K-1\\
\end{array} \right) -K\sum_{n=0}^{K-1}{\left( \begin{array}{c}
	K-1\\
	n\\
\end{array} \right) \left( -1 \right) ^n\left( \frac{1}{n+1} \right) ^{\mathsf{DOF}-m+1}}$, where $0\leqslant m\leqslant \mathsf{DOF}$ with $m\subseteq \mathbb{Z}$. Then, $y\left( K \right)=0$ when $m=\mathsf{DOF}$, and $y\left( K \right)>0$ when $0\leqslant m<\mathsf{DOF}$.

\textit{Proof}: When $m=\mathsf{DOF}$, we have
\begin{equation}
\label{eq71}
\begin{array}{l}
$$y\left( K \right) =1-\sum_{n=0}^{K-1}{\left( \begin{array}{c}
	K\\
	n+1\\
\end{array} \right) \left( -1 \right) ^n}=0$$.
\end{array}
\end{equation}
When $m<\mathsf{DOF}$, let $f\left( K \right) = \left( \begin{array}{c}
	\mathsf{DOF}-m+K-1\\
	K-1\\
\end{array} \right) $ and $g\left( K \right) = \sum_{n=0}^{K-1}{\left( \begin{array}{c}
	K-1\\
	n\\
\end{array} \right) \left( -1 \right) ^n\left( \frac{1}{n+1} \right) ^{\mathsf{DOF}-m+1}}$. It can be proved by mathematical induction that $f\left( K \right) \geqslant K$ for $K\geqslant 2$. Let $\alpha =\mathsf{DOF}-m+1$. We have
\begin{equation}
\label{eq72}
\begin{array}{l}
$$g\left( K \right) =\sum_{n=0}^{K-1}{\left( \begin{array}{c}
	K-1\\
	n\\
\end{array} \right) \left( -1 \right) ^n\frac{1}{\Gamma \left( \alpha \right)}\int_0^{\infty}{t^{\alpha -1}e^{-\left( n+1 \right) t}dt}}
\\
=\frac{1}{\Gamma \left( \alpha \right)}\int_0^{\infty}{t^{\alpha -1}e^{-t}}\left( 1-e^{-t} \right) ^{K-1}dt
\\
<\frac{1}{\Gamma \left( \alpha \right)}\int_0^{\infty}{t^{\alpha -1}e^{-t}}dt=1$$.
\end{array}
\end{equation}
Therefore, it follows that $y\left( K \right)>0$ when $0\leqslant m<\mathsf{DOF}$. This completes the proof. $\hfill\blacksquare$

Based on Lemma 6, it follows from (\ref{eq58}) and (\ref{eq70}) that the CAPA system has a lower array gain with multiple collaborative Eves than with multiple independent Eves for $K\geqslant 2$.
\vspace{-0.3cm}
\section{Simulation Results}
In this section, simulation results are presented to validate the accuracy of the approximations employed in the theoretical analyses and to evaluate the advantages of CAPA systems over the traditional SPDA systems in terms of secrecy rate and SOP. The simulation parameters are listed in Table \ref{tab1}, with $Q$ representing the number of terms taken from the series in (\ref{eq24}) to achieve an accurate approximation of $f_{\rho_{\mathrm{b}}}(x)$. Monte Carlo (MC) simulations with $2\times 10^6$ random channel realizations are employed to obtain the final results. The aperture spacing is set to half a wavelength, and the aperture length of each SPDA element is set to $\frac{\lambda}{5\sqrt{4\pi}}$.

\begin{figure}[htbp]%
    \centering
    \subfloat{
        \includegraphics[width=4cm]{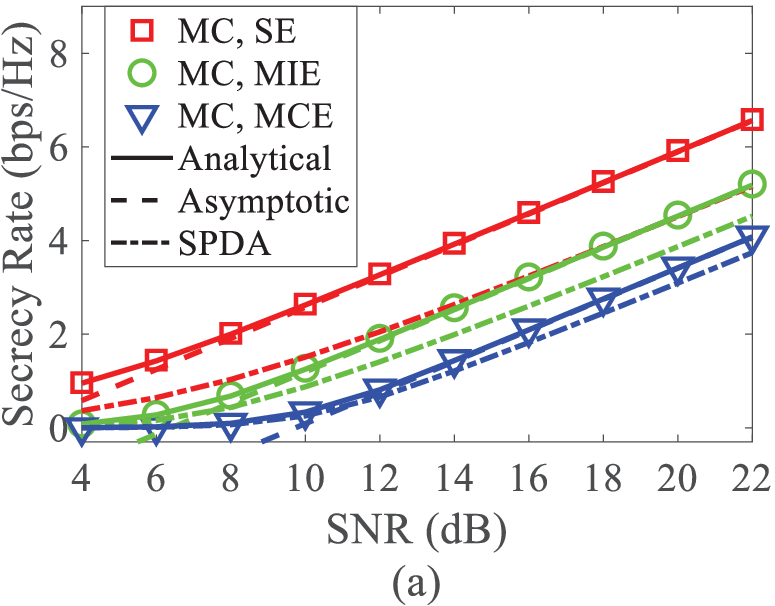}
        }\hfill
    \subfloat{
        \includegraphics[width=4cm]{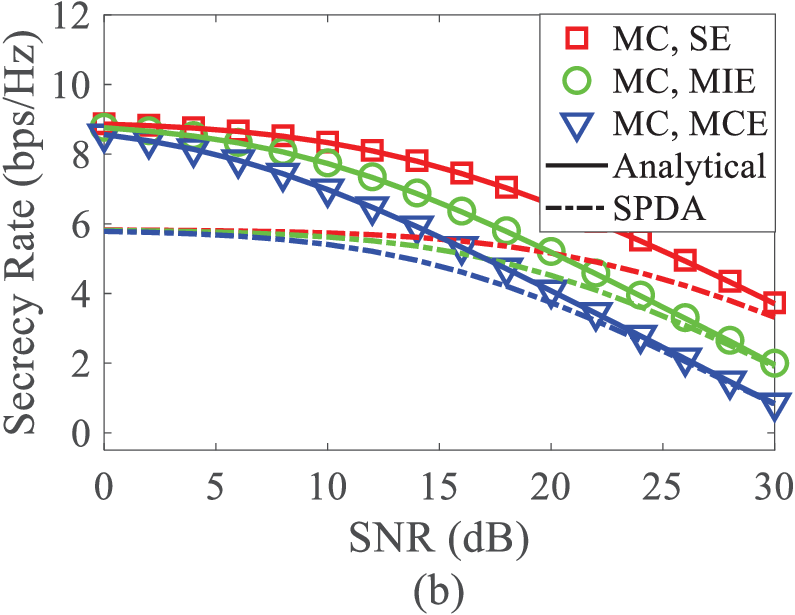}
        }\hfill
    \subfloat{
        \includegraphics[width=4cm]{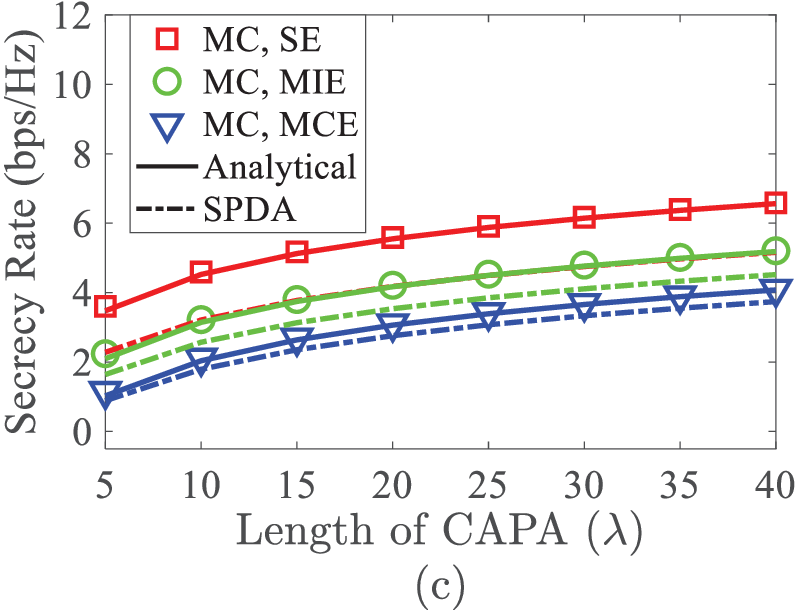}
        }\hfill
    \subfloat{
        \includegraphics[width=4cm]{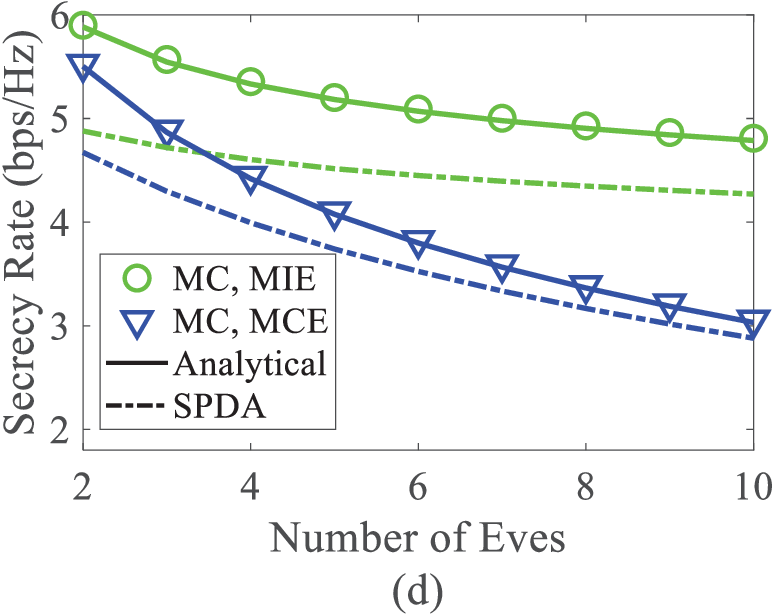}
        }
    \caption{Secrecy rate varying with (a) SNR of Bob, (b) SNR of Eve, (c) array length, and (d) the number of Eves.}
    \label{fig2}
    \vspace{-0.7cm}
\end{figure}
\vspace{-0.3cm}
\subsection{Secrecy Rate}
The secrecy rate as a function of Bob's SNR, Eve's SNR, array length, and the number of Eves is shown in Fig. \ref{fig2}(a), Fig. \ref{fig2}(b), Fig. \ref{fig2}(c), and Fig. \ref{fig2}(d), respectively. We use SE, MIE, and MCE to denote the scenarios of a single Eve, multiple independent Eves, and multiple collaborative Eves, respectively. As shown in Fig. \ref{fig2}(a), the analytical results closely match the MC simulations, demonstrating the accuracy of the approximation based on Landau's eigenvalue theorem. The analytical results also agree with the asymptotic results in the high-SNR region, demonstrating the correctness of the theoretical analyses. For a given SNR at Bob, the system with a single Eve achieves the highest secrecy rate, while the system with multiple collaborative Eves yields the lowest. The reason is that multiple collaborative Eves degrade the security performance more severely than a single Eve. Meanwhile, as shown in Fig. \ref{fig2}(a), the high-SNR slopes in all three scenarios are identical, while the CAPA system with a single Eve exhibits the smallest high-SNR offset and the CAPA system with multiple collaborative Eves has the largest high-SNR offset, which is consistent with the theoretical analyses in Section IV.

It can also be observed from Fig. \ref{fig2}(a) that, under the same scenario, the CAPA systems consistently outperform the SPDA systems in terms of secrecy rate. The reason is that the CAPA systems can better exploit EM characteristics and achieve higher secrecy rate. Meanwhile, the performance gain of the CAPA system over the SPDA system is larger in a single Eve scenario than in the multiple independent Eves and multiple collaborative Eves scenarios. This is because, in a single Eve scenario, the sidelobe suppression of the CAPA directly translates into improved secrecy performance. However, with multiple independent Eves or multiple collaborative Eves, the higher possibility of sidelobe leakage in the CAPA systems reduces its secrecy rate advantage compared with the SPDA systems. Moreover, the secrecy rates of the CAPA and SPDA systems are similar in the low-SNR region. This is because the potential of the CAPA system to utilize spatial resources has not been fully exploited in the low-SNR region. The advantages of the CAPA systems over the SPDA systems in terms of secrecy rate become more evident in the high-SNR region, where the spatial resources are fully utilized.

\begin{figure}[htbp]%
    \centering
    \subfloat{
        \includegraphics[width=4cm]{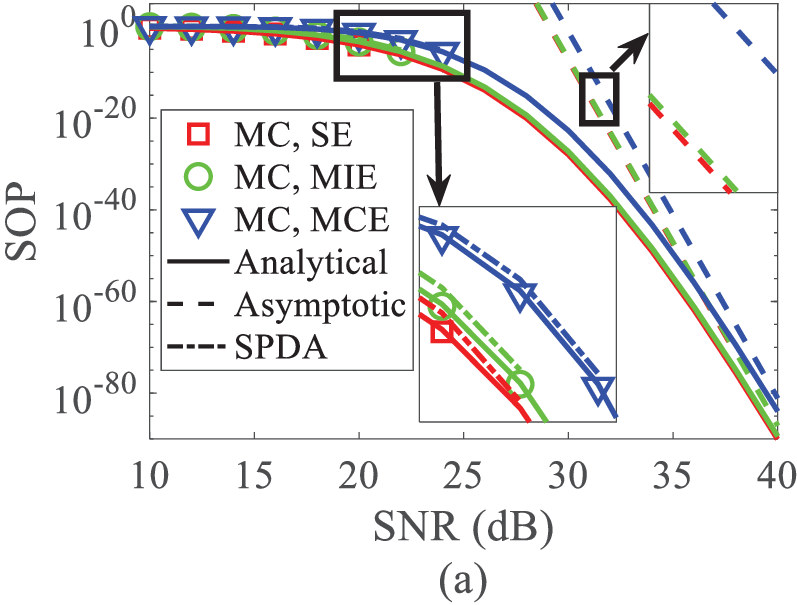}
        }\hfill
    \subfloat{
        \includegraphics[width=4cm]{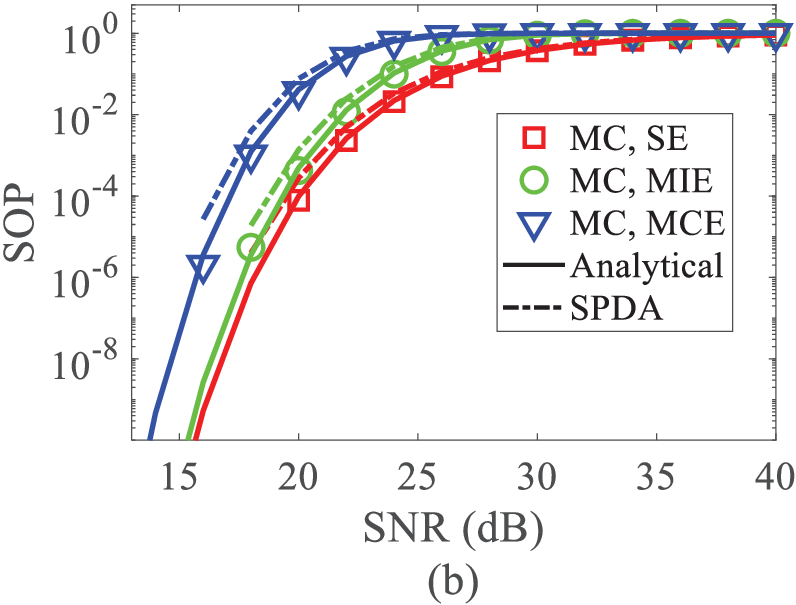}
        }\hfill
    \subfloat{
        \includegraphics[width=4cm]{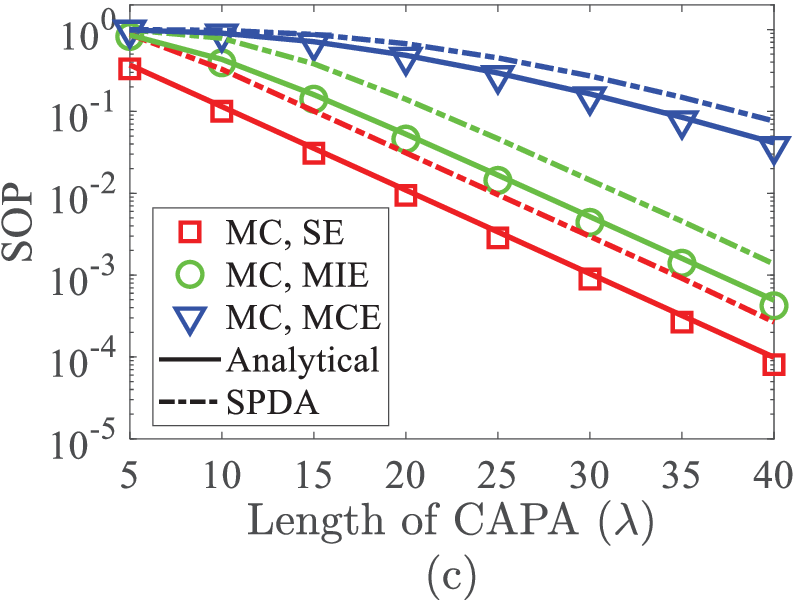}
        }\hfill
    \subfloat{
        \includegraphics[width=4cm]{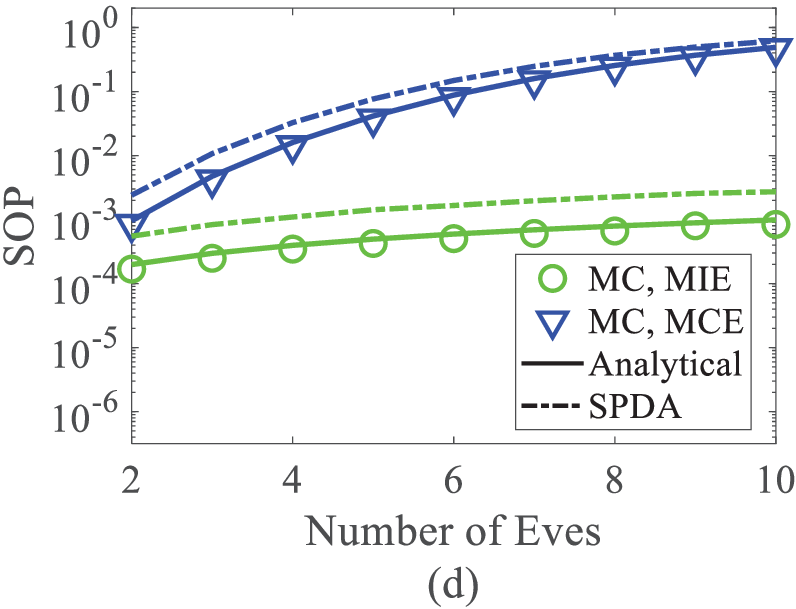}
        }
    \caption{SOP varying with (a) SNR of Bob, (b) SNR of Eve, (c) array length, and (d) the number of Eves.}
    \label{fig3}
    \vspace{-0.6cm}
\end{figure}
As shown in Fig. \ref{fig2}(b), the secrecy rate decreases as Eve's SNR increases in both the CAPA and SPDA systems. This is because the security performance of the wireless system would be more severely degraded when Eve's SNR is high, for example, when Eve is close to the legitimate users. It can also be observed that when Eves' SNR is low, the secrecy rates under the three scenarios are similar for both the CAPA and SPDA systems. The reason is that, at low Eve's SNR, the secrecy rate is primarily determined by Bob's SNR. Fig. \ref{fig2}(c) shows that increasing the array length leads to a corresponding increase in the secrecy rate for both systems. The reason is that the SNR of Bob's received signal improves more than that of the Eves' received signals due to MRT. Consequently, the corresponding mutual information and secrecy rate increase more than those of Eves. Meanwhile, the CAPA systems exhibit a greater secrecy rate improvement in a single Eve scenario than in the multiple independent Eves or multiple collaborative Eves scenarios, for reasons similar to those discussed in Fig. \ref{fig2}(a). Moreover, as illustrated in Fig. \ref{fig2}(d), the secrecy rates decrease with the number of Eves. It is also observed that the CAPA systems achieve a greater secrecy rate improvement in the multiple independent Eves scenario than in the multiple collaborative Eves scenario. This is because multiple collaborative Eves can exploit sidelobe leakage in the CAPA system, thereby reducing its secrecy advantage. As the number of Eves increases, the secrecy rate performance improvement of the CAPA systems over the SPDA systems decreases. The reason is similar to that discussed in Fig. \ref{fig2}(a).
\vspace{-0.4cm}
\subsection{SOP}
Fig. \ref{fig3}(a), Fig. \ref{fig3}(b), Fig. \ref{fig3}(c), and Fig. \ref{fig3}(d) illustrate SOP as a function of Bob's SNR, Eve's SNR, array length, and the number of Eves, respectively. As shown in Fig. \ref{fig3}(a), the analytical results agree with the MC simulations and are consistent with the asymptotic results in the high-SNR region, demonstrating the accuracy of the approximation based on Landau's eigenvalue theorem and the correctness of theoretical analyses. For a given SNR at Bob, the system with a single Eve exhibits the lowest SOP, while the system with multiple collaborative Eves shows the highest SOP, as multiple collaborative Eves degrade security more severely than a single Eve. Meanwhile, as shown in Fig. \ref{fig3}(a), the diversity orders in all three scenarios are equal to the spatial DoF, while the CAPA system with a single Eve exhibits the highest array gain, and the CAPA system with multiple collaborative Eves has the lowest array gain.

Fig. \ref{fig3}(b) shows that SOP increases with Eve's SNR for all systems, for reasons similar to those
discussed in Fig. 2(b). Similarly, as show in Fig. \ref{fig3}(c), increasing the array length leads to a decrease in SOP for both the CAPA and SPDA systems. Moreover, the CAPA systems achieve a greater SOP improvement in a single Eve scenario than in the multiple independent Eves or multiple collaborative Eves scenarios, for reasons similar to those discussed for the secrecy rate improvement in Fig. \ref{fig2}(c). Moreover, Fig. \ref{fig3}(d) indicates that SOP increases with the number of Eves, with the CAPA systems showing greater improvement in the multiple independent Eves scenario than in the multiple collaborative Eves scenario, following the same rationale as in Fig. \ref{fig2}(d).
\vspace{-0.3cm}
\section{Conclusion}
In this paper, we analyzed the secrecy rate and SOP of the CAPA systems under three scenarios: a single Eve, multiple independent Eves, and multiple collaborative Eves. Analytical expressions for secrecy rate and SOP were derived for all scenarios. The high-SNR slope, high-SNR power offset, diversity order, and array gain of the CAPA systems were obtained, with the diversity orders shown to equal the effective spatial DoF and the high-SNR slopes are the same in each case. Meanwhile, it is observed that the CAPA system with a single Eve has the smallest high-SNR offset and the highest array gain, whereas the CAPA system with multiple collaborative Eves has the largest high-SNR slope and the lowest array gain. The MC simulations were conducted, which agree with the analytical results and demonstrate the accuracy of the approximation based on Landau's eigenvalue theorem. For a given Bob's SNR, the CAPA system with a single Eve achieves the highest secrecy rate and the lowest SOP, whereas the system with multiple collaborative Eves exhibits the lowest secrecy rate and the highest SOP. Moreover, the secrecy rate of the CAPA systems increases, and SOP decreases, with higher Bob's SNR, longer array length, lower Eve's SNR, and fewer Eves. Furthermore, the CAPA system demonstrates better security performance in a single Eve scenario compared with the multiple independent Eves or multiple collaborative Eves scenarios.
\vspace{-0.4cm}
\begin{appendices}
\section{The Calculation of Eigenvalues $\sigma _{h_\mathrm{b}}^{2}\sigma _{\ell}$}
Based on (\ref{eq15}), for the $m$th eigenvalue of $R_{g_{\mathrm{b}}}(z,z^{\prime})$, where $m\subseteq \left\{ 1,2,...,N \right\}$, it follows that
\begin{equation}
\begin{small}
\label{eq73}
\begin{aligned}
&\int_{-\frac{L}{2}}^{\frac{L}{2}}{R_{g_{\mathrm{b}}}\left( z,z^{\prime} \right) \phi _m\left( z^{\prime} \right) dz^{\prime}}
\\
&=\sigma _{h_{\mathrm{b}}}^{2}\sum_{\ell =1}^N{\sigma _{\ell}\phi _{\ell}\left( z \right)}\int_{-\frac{L}{2}}^{\frac{L}{2}}{\phi _{\ell}^{*}\left( z^{\prime} \right) \phi _m\left( z^{\prime} \right) dz^{\prime}}
\\
&=\sigma _{h_{\mathrm{b}}}^{2}\sigma _m\phi _m\left( z \right).
\end{aligned}
\end{small}
\end{equation}
Meanwhile, using the Gauss-Legendre quadrature \cite{swarztrauber2003computing}, we obtain
\begin{equation}
\begin{small}
\label{eq74}
\int_{-\frac{L}{2}}^{\frac{L}{2}}{R_{g_{\mathrm{b}}}\left( z,z^{\prime} \right) \phi _m\left( z^{\prime} \right) dz^{\prime}}=\frac{L}{2}\sum_{k=1}^T{w_kR_{g_{\mathrm{b}}}\left( z,z_{k}^{\prime} \right) \phi _m\left( z_{k}^{\prime} \right) ,}
\end{small}
\end{equation}
where $z_{k}^{\prime}$ denotes the $k$th point and $w_k$ the corresponding weight. From (\ref{eq73}) and (\ref{eq74}), it follows that
\begin{equation}
\begin{small}
\label{eq75}
\frac{L}{2}\sum_{k=1}^T{w_kR_{g_{\mathrm{b}}}\left( z,z_{k}^{\prime} \right) \phi _m\left( z_{k}^{\prime} \right)}=\sigma _{h_{\mathrm{b}}}^{2}\sigma _m\phi _m\left( z \right) .
\end{small}
\end{equation}
Choosing the same points on $z$ abscissas, i.e., $z_a=z_{k}^{\prime}$ when $a=k$, we obtain
\begin{equation}
\begin{small}
\label{eq76}
\frac{L}{2}\sum_{k=1}^T{w_kR_{g_{\mathrm{b}}}\left( z_a,z_{k}^{\prime} \right) \phi _m\left( z_{k}^{\prime} \right)}=\sigma _{h_{\mathrm{b}}}^{2}\sigma _m\phi _m\left( z_a \right).
\end{small}
\end{equation}
For $\alpha \in \left\{ 1,...,T \right\} $ and $k\in \left\{ 1,...,T \right\}$, the expression can be written as shown in (\ref{eq77}) on the top of the next page. Consequently, the eigenvalue $\sigma _{h_\mathrm{b}}^{2}\sigma _{\ell}$ can be obtained by applying SVD to the matrix $\mathbf{F}$.

\begin{figure*}
\begin{small}
\begin{equation}
\label{eq77}
\underset{\mathbf{F}}{\underbrace{\frac{L}{2}\left( \begin{matrix}
	w_1R_{g_{\mathrm{b}}}\left( z_1,z_{1}^{\prime} \right)&		\cdots&		w_NR_{g_{\mathrm{b}}}\left( z_1,z_{N}^{\prime} \right)\\
	\cdots&		\cdots&		\cdots\\
	w_1R_{g_{\mathrm{b}}}\left( z_N,z_{1}^{\prime} \right)&		\cdots&		w_NR_{g_{\mathrm{b}}}\left( z_N,z_{N}^{\prime} \right)\\
\end{matrix} \right) }}\left( \begin{matrix}
	\phi _1\left( z_{1}^{\prime} \right)&		\cdots&		\phi _N\left( z_{1}^{\prime} \right)\\
	\cdots&		\cdots&		\cdots\\
	\phi _1\left( z_{N}^{\prime} \right)&		\cdots&		\phi _N\left( z_{N}^{\prime} \right)\\
\end{matrix} \right) =\sigma _{h_{\mathrm{b}}}^{2}\left( \begin{matrix}
	\sigma _1\phi _1\left( z_1 \right)&		\cdots&		\sigma _N\phi _N\left( z_1 \right)\\
	\cdots&		\cdots&		\cdots\\
	\sigma _1\phi _1\left( z_N \right)&		\cdots&		\sigma _N\phi _N\left( z_N \right)\\
\end{matrix} \right).
\end{equation}
\end{small}
\vspace{-0.8cm}
\end{figure*}

\vspace{-0.4cm}
\section{Secrecy Rate Under a Single Eve}
First, we introduce the Gamma function, $\Gamma \left( {n} \right)$, and the upper incomplete Gamma function, $\Gamma \left( {n,x} \right)$. These functions can be expressed as follows:
\begin{equation}
\begin{small}
\label{eq78}
\begin{aligned}
\Gamma \left( n \right) &=\int_0^{+\infty}{t^{n-1}e^{-t}dt}=\left\{ \begin{array}{c}
	\mathrm{E}_1\left( 0 \right) ,n=0\\
	\left( n-1 \right) !,n\in \mathbb{Z} ^+\\
\end{array} \right. ,
\end{aligned}
\end{small}
\end{equation}
and
\begin{equation}
\begin{small}
\label{eq79}
\begin{aligned}
\Gamma \left( n,x \right) &=\int_x^{+\infty}{t^{n-1}e^{-t}dt}
\\
&=\left\{ \begin{array}{c}
	\mathrm{E}_1\left( x \right) ,n=0\\
	\left( n-1 \right) !e^{-x}\sum_{k=0}^{n-1}{\frac{x^k}{k!},n\in \mathbb{Z} ^+.}\\
\end{array} \right.
\end{aligned}
\end{small}
\end{equation}
where $\mathrm{E}_1\left( x \right) $ denotes the exponential integral of order 1 and can be expressed as follows:
\begin{equation}
\label{eq80}
\begin{small}
\mathrm{E}_1\left( x \right) =\int_x^{\infty}{\frac{e^{-u}}{u}du}.
\end{small}
\end{equation}

Meanwhile, the function $\mathrm{F}\left( n+1,x \right)$ is defined as follows:
\begin{equation}
\begin{small}
\label{eq81}
\begin{array}{l}
{\rm{F}}\left( {n + 1,x} \right) = \int_x^{ + \infty } {\ln u \cdot {u^n} \cdot {e^{ - u}}du} \\
 = n!\left( {\ln x \cdot {e^{ - x}} + {{\rm{E}}_1}\left( x \right) + \sum\limits_{k = 0}^{n - 1} {\frac{{\ln x \cdot {x^{n - k}} \cdot {e^{ - x}} + \Gamma \left( {n - k,x} \right)}}{{\left( {n - k} \right)!}}} } \right)\\
,n \in {\rm{N}}.
\end{array}
\end{small}
\end{equation}
When $n=1$, we have $\mathrm{F}\left( 1,x \right) =\ln x\cdot e^{-x}+\mathrm{E}_1\left( x \right) $. Moreover, the following result is used
\begin{equation}
\begin{small}
\label{eq82}
\int_0^x{u^ne^{-u}du}=n!\left( 1-e^{-x}\sum_{k=0}^n{\frac{x^k}{k!}} \right) ,n\in \mathrm{N}.
\end{small}
\end{equation}

Then, based on (\ref{eq41}), the secrecy rate of the CAPA system can be expressed as shown in (\ref{eq83}) on the top of the next page.
\begin{figure*}
\begin{small}
\begin{equation}
\label{eq83}
\begin{array}{l}
R_\mathrm{S}\approx \frac{1}{\overline{\gamma }_\mathrm{b}\overline{\gamma }_\mathrm{e}\prod_{\ell =1}^{\mathsf{DOF}}{\sigma _{\ell}}}\sum_{q=0}^{\infty}{\frac{\psi _q}{\sigma _{\min}^{q}\Gamma (\mathsf{DOF}+q)}}\left\{ \left( \frac{1}{\overline{\gamma }_\mathrm{b}} \right) ^{\mathsf{DOF}+q-1}e^{\frac{1}{\overline{\gamma }_\mathrm{b}\sigma _{\min}}}\sum_{k=0}^{\mathsf{DOF}+q-1}{\left( \begin{array}{c}
	\mathsf{DOF}+q-1\\
	k\\
\end{array} \right) \left( -1 \right) ^{\mathsf{DOF}+q-1-k}} \right.
\\
\cdot\left( \overline{\gamma }_\mathrm{b}\sigma _{\min} \right) ^{k+1}\left[ \frac{k!}{\ln 2}\left( \frac{\overline{\gamma }_\mathrm{b}\overline{\gamma }_\mathrm{e}\sigma _{\min}}{\overline{\gamma }_\mathrm{b}\sigma _{\min}+\overline{\gamma }_\mathrm{e}}e^{\frac{1}{\overline{\gamma }_\mathrm{e}}}\left( e^{-\frac{\overline{\gamma }_\mathrm{b}\sigma _{\min}+\overline{\gamma }_\mathrm{e}}{\overline{\gamma }_\mathrm{b}\overline{\gamma }_\mathrm{e}\sigma _{\min}}}\ln \frac{1}{\overline{\gamma }_\mathrm{b}\sigma _{\min}}+\mathrm{E}_1\left( \frac{\overline{\gamma }_\mathrm{b}\sigma _{\min}+\overline{\gamma }_\mathrm{e}}{\overline{\gamma }_\mathrm{b}\overline{\gamma }_\mathrm{e}\sigma _{\min}} \right) \right) \right. \right.
\\
+\overline{\gamma }_\mathrm{e}\left( \mathrm{E}_1\left( \frac{1}{\overline{\gamma }_\mathrm{b}\sigma _{\min}} \right) -e^{\frac{1}{\overline{\gamma }_\mathrm{e}}}\mathrm{E}_1\left( \frac{\overline{\gamma }_\mathrm{b}\sigma _{\min}+\overline{\gamma }_\mathrm{e}}{\overline{\gamma }_\mathrm{b}\overline{\gamma }_\mathrm{e}\sigma _{\min}} \right) \right) +\sum_{p=0}^{k-1}{\frac{1}{\left( k-p \right) !}\left( \overline{\gamma }_\mathrm{b}\sigma _{\min}e^{\frac{1}{\gamma _\mathrm{e}}}\left( \frac{\overline{\gamma }_\mathrm{e}}{\overline{\gamma }_\mathrm{b}\sigma _{\min}+\overline{\gamma }_\mathrm{e}} \right) ^{k-p+1} \right.}
\\
\cdot\left( \mathrm{F}\left( k-p+1,\frac{\overline{\gamma }_\mathrm{b}\sigma _{\min}+\overline{\gamma }_\mathrm{e}}{\overline{\gamma }_\mathrm{b}\overline{\gamma }_\mathrm{e}\sigma _{\min}} \right) +\Gamma \left( k-p+1,\frac{\overline{\gamma }_\mathrm{b}\sigma _{\min}+\overline{\gamma }_\mathrm{e}}{\overline{\gamma }_\mathrm{b}\overline{\gamma }_\mathrm{e}\sigma _{\min}} \right) \ln \frac{\overline{\gamma }_\mathrm{e}}{\overline{\gamma }_\mathrm{b}\sigma _{\min}+\overline{\gamma }_\mathrm{e}} \right)
\\
\left. \left. +\left( k-p-1 \right) !e^{\frac{1}{\overline{\gamma }_\mathrm{e}}}\overline{\gamma }_\mathrm{b}\sigma _{\min}\sum_{r=0}^{k-p-1}{\frac{\left( \overline{\gamma }_\mathrm{e} \right) ^{r+1}}{\left( \overline{\gamma }_\mathrm{b}\sigma _{\min}+\overline{\gamma }_\mathrm{e} \right) ^{r+1}r!}\Gamma \left( r+1,\frac{\overline{\gamma }_\mathrm{b}\sigma _{\min}+\overline{\gamma }_\mathrm{e}}{\overline{\gamma }_\mathrm{b}\overline{\gamma }_\mathrm{e}\sigma _{\min}} \right)} \right) \right)
\\
\left. +\log _2\left( \overline{\gamma }_\mathrm{b}\sigma _{\min} \right) \cdot k!e^{\frac{1}{\overline{\gamma }_\mathrm{e}}}\overline{\gamma }_\mathrm{b}\sigma _{\min}\sum_{n=0}^k{\frac{\left( \overline{\gamma }_\mathrm{e} \right) ^{n+1}}{\left( \overline{\gamma }_\mathrm{b}\sigma _{\min}+\overline{\gamma }_\mathrm{e} \right) ^{n+1}n!}\Gamma \left( n+1,\frac{\overline{\gamma }_\mathrm{b}\sigma _{\min}+\overline{\gamma }_\mathrm{e}}{\overline{\gamma }_\mathrm{b}\overline{\gamma }_\mathrm{e}\sigma _{\min}} \right)} \right]
\\
-\frac{\overline{\gamma }_\mathrm{b}\sigma _{\min}^{\mathsf{DOF}+q}\cdot \left( \mathsf{DOF}+q-1 \right) !e^{\frac{\overline{\gamma }_\mathrm{b}\sigma _{\min}+\overline{\gamma }_\mathrm{e}}{\overline{\gamma }_\mathrm{b}\overline{\gamma }_\mathrm{e}\sigma _{\min}}}}{\ln 2}\sum_{m=0}^{\mathsf{DOF}+q-1}{\frac{\left( \frac{1}{\overline{\gamma }_\mathrm{b}\sigma _{\min}} \right) ^m}{m!}}\sum_{t=0}^m{\left( \begin{array}{c}
	m\\
	t\\
\end{array} \right) \left( -1 \right) ^{m-t}}\left( \frac{\overline{\gamma }_\mathrm{b}\overline{\gamma }_\mathrm{e}\sigma _{\min}}{\overline{\gamma }_\mathrm{b}\sigma _{\min}+\overline{\gamma }_\mathrm{e}} \right) ^{t+1}
\\
\cdot\left. \left[ \mathrm{F}\left( t+1,\frac{\overline{\gamma }_\mathrm{b}\sigma _{\min}+\overline{\gamma }_\mathrm{e}}{\overline{\gamma }_\mathrm{b}\overline{\gamma }_\mathrm{e}\sigma _{\min}} \right) +\Gamma \left( t+1,\frac{\overline{\gamma }_\mathrm{b}\sigma _{\min}+\overline{\gamma }_\mathrm{e}}{\overline{\gamma }_\mathrm{b}\overline{\gamma }_\mathrm{e}\sigma _{\min}} \right) \ln \frac{\overline{\gamma }_\mathrm{b}\overline{\gamma }_\mathrm{e}\sigma _{\min}}{\overline{\gamma }_\mathrm{b}\sigma _{\min}+\overline{\gamma }_\mathrm{e}} \right] \right\}.
\end{array}
\end{equation}
\vspace{-0.6cm}
\end{small}
\end{figure*}
\vspace{-0.4cm}
\section{Diversity Order and Array Gain Under a Single Eve}
For the sake of conciseness, we denote $1/\overline{\gamma }_{\mathrm{b}}$ and $\mathsf{DOF}$ as $z$ and $N$, respectively. The expression of SOP under the target secrecy rate $R_0$ and Bob's high-SNR region can be approximated as follows:

\begin{equation}
\begin{small}
\label{eq84}
\begin{array}{l}
P_{\mathrm{SOP}_{\mathrm{S}}}\left( R_0 \right) \approx \frac{\sigma _{\min}^{\mathsf{DOF}}}{\prod_{\ell =1}^{\mathsf{DOF}}{\sigma _{\ell}}}\sum_{q=0}^{\infty}{\psi _q\left( 1-1/\underset{A}{\underbrace{\left( 1+\frac{2^{R_0}\overline{\gamma }_{\mathrm{e}}}{\sigma _{\min}}z \right) ^{N+q}}} \right.}
\\
\cdot \underset{B}{\underbrace{\left( \sum_{n=0}^N{\frac{\left( -\frac{2^{R_0}-1}{\sigma _{\min}} \right) ^n}{n!}z^n}+o\left( z^N \right) \right) }}\underset{C}{\underbrace{\sum_{k=0}^{N+q-1}{\frac{1}{\sigma _{\min}^{k}}z^k}}}
\\
\left. \underset{C}{\underbrace{\cdot \sum_{m=0}^k{\frac{\left( 2^{R_0}-1 \right) ^{k-m}\left( 2^{R_0}\overline{\gamma }_{\mathrm{e}} \right) ^m}{\left( k-m \right) !}\left( \frac{\sigma _{\min}}{\sigma _{\min}+2^{R_0}\overline{\gamma }_{\mathrm{e}}z} \right) ^{m+1}}}} \right) .
\end{array}
\end{small}
\end{equation}

Then, we will then show that the coefficient of $z^M$ in (\ref{eq84}) is zero for $0\leqslant M\leqslant N-1$ and non-zero for $M=N$. The coefficient of $z^k$, for $k\in \mathbb{Z} ^+,0\leqslant k\leqslant N-1$, in the term $C$ can be expressed as follows:
\begin{equation}
\begin{small}
\label{eq85}
\begin{aligned}
\mathrm{coeff}\left( z_{C}^{k} \right) =&\left( \frac{1}{\sigma _{\min}} \right) ^k\sum_{a=0}^k{\sum_{b=0}^a{\left( \begin{array}{c}
	N+q-k-1+a\\
	a-b\\
\end{array} \right)}}
\\
&\cdot \frac{\left( 2^{R_0}-1 \right) ^b\left( 2^{R_0}\overline{\gamma }_{\mathrm{e}} \right) ^{k-b}}{b!}.
\end{aligned}
\end{small}
\end{equation}

Subsequently, the coefficient of $z^M$, for $M\in \mathbb{Z} ^+$ and $k\leqslant M\leqslant N-1$, in the term $ B*C$ can be expressed as follows:
\begin{equation}
\begin{small}
\label{eq86}
\begin{array}{l}
\mathrm{coeff}\left( z_{B*C}^{M} \right) =\sum_{k=0}^M{\frac{\left( -\frac{2^{R_0}-1}{\sigma _{\min}} \right) ^{M-k}}{\left( M-k \right) !}\left( \frac{1}{\sigma _{\min}} \right) ^k\sum_{a=0}^k{\sum_{b=0}^a{}}}
\\
\cdot \left( \begin{array}{c}
	N+q-k-1+a\\
	a-b\\
\end{array} \right) \frac{\left( 2^{R_0}-1 \right) ^b\left( 2^{R_0}\overline{\gamma }_{\mathrm{e}} \right) ^{k-b}}{b!}.
\end{array}
\end{small}
\end{equation}

Next, the coefficient of $z^M$, for $M\in \mathbb{Z} ^+$ and $k\leqslant M\leqslant N-1$, in the term $A$ is
\begin{equation}
\begin{small}
\label{eq87}
\begin{array}{l}
\mathrm{coeff}\left( z_{A}^{M} \right) =\left( \begin{array}{c}
	N+q\\
	M\\
\end{array} \right) \left( \frac{2^{R_0}\overline{\gamma }_{\mathrm{e}}}{\sigma _{\min}} \right) ^M.
\end{array}
\end{small}
\end{equation}

Then, it can be shown that $\mathrm{coeff}\left( z_{B*C}^{M} \right) = \mathrm{coeff}\left( z_{A}^{M} \right)$, as given in (\ref{eq88}) on the top of the next page. Hence, the coefficient of $z^M$ in (\ref{eq84}) is zero for $0\leqslant M\leqslant N-1$.

\begin{figure*}
\begin{equation}
\label{eq88}
\begin{small}
\begin{array}{l}
\mathrm{coeff}\left( z_{B*C}^{M} \right) =\sum_{k=0}^M{\frac{\left( -\frac{2^{R_0}-1}{\sigma _{\min}} \right) ^{M-k}}{\left( M-k \right) !}\left( \frac{1}{\sigma _{\min}} \right) ^k\sum_{b=0}^k{\sum_{a=b}^k{\left( \begin{array}{c}
	N+q-k-1+a\\
	a-b\\
\end{array} \right) \frac{\left( 2^{R_0}-1 \right) ^b\left( 2^{R_0}\overline{\gamma }_{\mathrm{e}} \right) ^{k-b}}{b!}}}}
\\
=\sum_{b=0}^M{\sum_{k=b}^M{\frac{\left( -\frac{2^{R_0}-1}{\sigma _{\min}} \right) ^{M-k}}{\left( M-k \right) !}\left( \frac{1}{\sigma _{\min}} \right) ^k\frac{\left( 2^{R_0}-1 \right) ^b\left( 2^{R_0}\overline{\gamma }_{\mathrm{e}} \right) ^{k-b}}{b!}\left( \begin{array}{c}
	N+q\\
	k-b\\
\end{array} \right)}}=\left( \begin{array}{c}
	N+q\\
	M\\
\end{array} \right) \left( \frac{2^{R_0}\overline{\gamma }_{\mathrm{e}}}{\sigma _{\min}} \right) ^M=\mathrm{coeff}\left( z_{A}^{M} \right).
\end{array}
\end{small}
\end{equation}
\vspace{-0.8cm}
\end{figure*}

When $M=N$ and $q\geqslant 1$ hold, using the similar approach, it can be found that the coefficient of $z^N$ in (\ref{eq84}) is zero. When $M=N$ and $q = 0$ hold, the coefficient of $z^N$ in the term $C$ can be expressed as follows:
\begin{equation}
\begin{small}
\label{eq89}
\begin{aligned}
\mathrm{coeff}\left( z_{C}^{N} \right) =&\left( \frac{1}{\sigma _{\min}} \right) ^N\sum_{a=1}^{N-1}{\sum_{b=1}^a{\left( \begin{array}{c}
	a\\
	a-b+1\\
\end{array} \right)}}
\\
&\cdot \frac{\left( 2^{R_0}-1 \right) ^b\left( 2^{R_0}\overline{\gamma }_{\mathrm{e}} \right) ^{N-b}}{b!}.
\end{aligned}
\end{small}
\end{equation}
Therefore, the coefficient of $z^N$ in the term $B*C$ is
\begin{small}
\begin{equation}
\label{eq90}
\begin{array}{l}
\mathrm{coeff}\left( z_{B*C}^{N} \right) =\underset{D}{\underbrace{\frac{\left( -\frac{2^{R_0}-1}{\sigma _{\min}} \right) ^0}{0!}\left( \frac{1}{\sigma _{\min}} \right) ^N\sum_{a=1}^{N-1}{\sum_{b=1}^a{}}}}
\\
\underset{D}{\underbrace{\cdot \left( \begin{array}{c}
	a\\
	a-b+1\\
\end{array} \right) \frac{\left( 2^{R_0}-1 \right) ^b\left( 2^{R_0}\overline{\gamma }_{\mathrm{e}} \right) ^{N-b}}{b!}}}
\\
+\underset{E}{\underbrace{\sum_{k=0}^{N-1}{\frac{\left( -\frac{2^{R_0}-1}{\sigma _{\min}} \right) ^{N-k}}{\left( N-k \right) !\left( \sigma _{\min} \right) ^k}\sum_{a=0}^k{\sum_{b=0}^a{}}}}}
\\
\underset{E}{\underbrace{\cdot \left( \begin{array}{c}
	N-k-1+a\\
	a-b\\
\end{array} \right) \frac{\left( 2^{R_0}-1 \right) ^b\left( 2^{R_0}\overline{\gamma }_{\mathrm{e}} \right) ^{k-b}}{b!}}}.
\end{array}
\end{equation}
\end{small}

In the term $D$, we obtain
\begin{equation}
\begin{small}
\label{eq91}
\begin{aligned}
\mathrm{coeff}\left( z_{D}^{N} \right) =&\left( \frac{2^{R_0}\overline{\gamma }_{\mathrm{e}}}{\sigma _{\min}} \right) ^N\sum_{m=1}^{N-1}{\left( \frac{2^{R_0}-1}{2^{R_0}\overline{\gamma }_{\mathrm{e}}} \right) ^{N-m}}
\\
&\cdot \frac{1}{\left( N-m \right) !}\sum_{c=0}^{m-1}{\left( \begin{array}{c}
	N-m+c\\
	c+1\\
\end{array} \right)}.
\end{aligned}
\end{small}
\end{equation}
In the term $E$, it follows that
\begin{small}
\begin{equation}
\label{eq92}
\begin{small}
\begin{array}{l}
$$\mathrm{coeff}\left( z_{E}^{N} \right) =-\left( \frac{2^{R_0}-1}{\sigma _{\min}} \right) ^N\sum_{m=0}^{N-1}{\left( \begin{array}{c}
	N\\
	m\\
\end{array} \right)}\left( \frac{2^{R_0}\overline{\gamma }_{\mathrm{e}}}{2^{R_0}-1} \right) ^m\frac{1}{\left( N-m \right) !}$$.
\end{array}
\end{small}
\end{equation}
\end{small}
Based on (\ref{eq91}) and (\ref{eq92}), we can get the result in (\ref{eq90}) as follows:
\begin{equation}
\begin{small}
\label{eq93}
\begin{array}{l}
$$\mathrm{coeff}\left( z_{B*C}^{N} \right) =-\left( \frac{2^{R_0}\overline{\gamma }_{\mathrm{e}}}{\sigma _{\min}} \right) ^N\sum_{m=0}^{N-1}{\left( \frac{2^{R_0}-1}{2^{R_0}\overline{\gamma }_{\mathrm{e}}} \right) ^{N-m}\frac{1}{\left( N-m \right) !}}$$.
\end{array}
\end{small}
\end{equation}
Thus, under Bob's high-SNR region, the coefficient of $z^N$ in (\ref{eq84}) is
\begin{equation}
\begin{small}
\label{eq94}
\begin{array}{l}
$$\mathrm{coeff}\left( z^N \right) =\frac{\sigma _{\min}^{N}}{\prod_{\ell =1}^{N}{\sigma _{\ell}}}\psi _0\left[ \left( \frac{2^{R_0}\overline{\gamma }_{\mathrm{e}}}{\sigma _{\min}} \right) ^N-\mathrm{coeff}\left( z_{B*C}^{N} \right) \right]
\\
=\frac{1}{\prod_{\ell =1}^{N}{\sigma _{\ell}}}\psi _0\sum_{m=0}^N{\left( \frac{2^{R_0}-1}{2^{R_0}\overline{\gamma }_{\mathrm{e}}} \right) ^m\frac{1}{m!}\left( 2^{R_0}\overline{\gamma }_{\mathrm{e}} \right) ^N}>0$$.
\end{array}
\end{small}
\end{equation}

Therefore, in Bob's high-SNR region, (\ref{eq46}) can be rewritten as follows:
\begin{equation}
\begin{small}
\label{eq95}
\begin{array}{l}
$$\underset{\overline{\gamma }_{\mathrm{b}}\rightarrow \infty}{\lim}P_{\mathrm{SOP}_{\mathrm{S}}}\left( R_0 \right) \approx \left\{ \left[ \left( \prod_{\ell =1}^{\mathsf{DOF}}{\sigma _{\ell}} \right) \right. / \right.
\\
\left. \left. \left( \psi _0\sum_{m=0}^{\mathsf{DOF}}{\frac{1}{m!}\left( \frac{2^{R_0}-1}{2^{R_0}\overline{\gamma }_{\mathrm{e}}} \right) ^m} \right) \right] ^{\frac{1}{\mathsf{DOF}}}\frac{\overline{\gamma }_{\mathrm{b}}}{2^{R_0}\overline{\gamma }_{\mathrm{e}}} \right\} ^{-\mathsf{DOF}}$$.
\end{array}
\end{small}
\end{equation}

Accordingly, the diversity order is $\mathsf{DOF}$, and the array gain is given in (\ref{eq47}).

$\hfill\blacksquare$
\vspace{-0.4cm}
\section{Secrecy Rate Under Multiple Independent Eves}
Based on (\ref{eq50}), the secrecy rate of the CAPA system under multiple independent Eves is given in (\ref{eq96}) on the top of the next page.
\begin{figure*}
\begin{small}
\begin{equation}
\label{eq96}
\begin{array}{l}
R_\mathrm{M}\approx K\sum_{a=0}^{K-1}{\left( \begin{array}{c}
	K-1\\
	a\\
\end{array} \right)}\frac{\left( -1 \right) ^a}{\overline{\gamma }_\mathrm{b}\overline{\gamma }_\mathrm{e}\prod_{\ell =1}^{\mathsf{DOF}}{\sigma _{\ell}}}\sum_{q=0}^{\infty}{\frac{\psi _q}{\sigma _{\min}^{q}\Gamma (\mathsf{DOF}+q)}}\left\{ \left( \frac{1}{\overline{\gamma }_\mathrm{b}} \right) ^{\mathsf{DOF}+q-1}e^{\frac{1}{\overline{\gamma }_\mathrm{b}\sigma _{\min}}}\sum_{k=0}^{\mathsf{DOF}+q-1}{\left( \begin{array}{c}
	\mathsf{DOF}+q-1\\
	k\\
\end{array} \right)} \right.
\\
\cdot\left( -1 \right) ^{\mathsf{DOF}+q-1-k}\left( \overline{\gamma }_\mathrm{b}\sigma _{\min} \right) ^{k+1}\left[ \frac{k!}{\ln 2}\left( \frac{\overline{\gamma }_\mathrm{b}\overline{\gamma }_\mathrm{e}\sigma _{\min}}{\left( 1+a \right) \overline{\gamma }_\mathrm{b}\sigma _{\min}+\overline{\gamma }_\mathrm{e}}e^{\frac{1+a}{\overline{\gamma }_\mathrm{e}}}\left( e^{-\frac{\left( 1+a \right) \overline{\gamma }_\mathrm{b}\sigma _{\min}+\overline{\gamma }_\mathrm{e}}{\overline{\gamma }_\mathrm{b}\overline{\gamma }_\mathrm{e}\sigma _{\min}}}\ln \frac{1}{\overline{\gamma }_\mathrm{b}\sigma _{\min}}+\mathrm{E}_1\left( \frac{\left( 1+a \right) \overline{\gamma }_\mathrm{b}\sigma _{\min}+\overline{\gamma }_\mathrm{e}}{\overline{\gamma }_\mathrm{b}\overline{\gamma }_\mathrm{e}\sigma _{\min}} \right) \right) \right. \right.
\\
+\frac{\overline{\gamma }_\mathrm{e}}{1+a}\left( \mathrm{E}_1\left( \frac{1}{\overline{\gamma }_\mathrm{b}\sigma _{\min}} \right) -e^{\frac{1+a}{\overline{\gamma }_\mathrm{e}}}\mathrm{E}_1\left( \frac{\left( 1+a \right) \overline{\gamma }_\mathrm{b}\sigma _{\min}+\overline{\gamma }_\mathrm{e}}{\overline{\gamma }_\mathrm{b}\overline{\gamma }_\mathrm{e}\sigma _{\min}} \right) \right) +\sum_{p=0}^{k-1}{\frac{1}{\left( k-p \right) !}\left( \overline{\gamma }_\mathrm{b}\sigma _{\min}e^{\frac{1+a}{\overline{\gamma }_\mathrm{e}}}\left( \frac{\overline{\gamma }_\mathrm{e}}{\left( 1+a \right) \overline{\gamma }_\mathrm{b}\sigma _{\min}+\overline{\gamma }_\mathrm{e}} \right) ^{k-p+1} \right.}
\\
\cdot\left( \mathrm{F}\left( k-p+1,\frac{\left( 1+a \right) \overline{\gamma }_\mathrm{b}\sigma _{\min}+\overline{\gamma }_\mathrm{e}}{\overline{\gamma }_\mathrm{b}\overline{\gamma }_\mathrm{e}\sigma _{\min}} \right) +\Gamma \left( k-p+1,\frac{\left( 1+a \right) \overline{\gamma }_\mathrm{b}\sigma _{\min}+\overline{\gamma }_\mathrm{e}}{\overline{\gamma }_\mathrm{b}\overline{\gamma }_\mathrm{e}\sigma _{\min}} \right) \ln \frac{\overline{\gamma }_\mathrm{e}}{\left( 1+a \right) \overline{\gamma }_\mathrm{b}\sigma _{\min}+\overline{\gamma }_\mathrm{e}} \right)
\\
\left. \left. +\left( k-p-1 \right) !e^{\frac{1+a}{\overline{\gamma }_\mathrm{e}}}\overline{\gamma }_\mathrm{b}\sigma _{\min}\sum_{r=0}^{k-p-1}{\frac{\left( \overline{\gamma }_\mathrm{e} \right) ^{r+1}}{\left( \left( 1+a \right) \overline{\gamma }_\mathrm{b}\sigma _{\min}+\overline{\gamma }_\mathrm{e} \right) ^{r+1}r!}\Gamma \left( r+1,\frac{\left( 1+a \right) \overline{\gamma }_\mathrm{b}\sigma _{\min}+\overline{\gamma }_\mathrm{e}}{\overline{\gamma }_\mathrm{b}\overline{\gamma }_\mathrm{e}\sigma _{\min}} \right)} \right) \right)
\\
\left. +\log _2\left( \overline{\gamma }_\mathrm{b}\sigma _{\min} \right) \cdot k!e^{\frac{1+a}{\overline{\gamma }_\mathrm{e}}}\overline{\gamma }_\mathrm{b}\sigma _{\min}\sum_{n=0}^k{\frac{\left( \overline{\gamma }_\mathrm{e} \right) ^{n+1}}{\left( \left( 1+a \right) \overline{\gamma }_\mathrm{b}\sigma _{\min}+\overline{\gamma }_\mathrm{e} \right) ^{n+1}n!}\Gamma \left( n+1,\frac{\left( 1+a \right) \overline{\gamma }_\mathrm{b}\sigma _{\min}+\overline{\gamma }_\mathrm{e}}{\overline{\gamma }_\mathrm{b}\overline{\gamma }_\mathrm{e}\sigma _{\min}} \right)} \right]
\\
-\frac{\overline{\gamma }_\mathrm{b}\sigma _{\min}^{\mathsf{DOF}+q}\cdot \left( \mathsf{DOF}+q-1 \right) !e^{\frac{\left( 1+a \right) \overline{\gamma }_\mathrm{b}\sigma _{\min}+\overline{\gamma }_\mathrm{e}}{\overline{\gamma }_\mathrm{b}\overline{\gamma }_\mathrm{e}\sigma _{\min}}}}{\ln 2}\sum_{m=0}^{\mathsf{DOF}+q-1}{\frac{\left( \frac{1}{\overline{\gamma }_\mathrm{b}\sigma _{\min}} \right) ^m}{m!}}\sum_{t=0}^m{\left( \begin{array}{c}
	m\\
	t\\
\end{array} \right) \left( -1 \right) ^{m-t}}\left( \frac{\overline{\gamma }_\mathrm{b}\overline{\gamma }_\mathrm{e}\sigma _{\min}}{\left( 1+a \right) \overline{\gamma }_\mathrm{b}\sigma _{\min}+\overline{\gamma }_\mathrm{e}} \right) ^{t+1}
\\
\cdot\left. \left[ \mathrm{F}\left( t+1,\frac{\left( 1+a \right) \overline{\gamma }_\mathrm{b}\sigma _{\min}+\overline{\gamma }_\mathrm{e}}{\overline{\gamma }_\mathrm{b}\overline{\gamma }_\mathrm{e}\sigma _{\min}} \right) +\Gamma \left( t+1,\frac{\left( 1+a \right) \overline{\gamma }_\mathrm{b}\sigma _{\min}+\overline{\gamma }_\mathrm{e}}{\overline{\gamma }_\mathrm{b}\overline{\gamma }_\mathrm{e}\sigma _{\min}} \right) \ln \frac{\overline{\gamma }_\mathrm{b}\overline{\gamma }_\mathrm{e}\sigma _{\min}}{\left( 1+a \right) \overline{\gamma }_\mathrm{b}\sigma _{\min}+\overline{\gamma }_\mathrm{e}} \right] \right\}.
\end{array}
\end{equation}
\vspace{-0.8cm}
\end{small}
\end{figure*}

\vspace{-0.4cm}
\section{Diversity Order and Array Gain Under Multiple Independent Eves}
For the sake of conciseness, $1/\overline{\gamma }_{\mathrm{b}}$ and $\mathsf{DOF}$ are denoted as $z$ and $N$, respectively. Using a proof similar to Theorem 1, it can be found that in Bob's high-SNR region, the terms $z^k$, for $k\in \mathbb{Z} ^+$ and $0\leqslant k\leqslant N-1$, in (\ref{eq57}) cancel out. The coefficient of $z^N$ is
\begin{equation}
\begin{small}
\label{eq97}
\begin{aligned}
\mathrm{coeff}\left( z^N \right) &=\frac{K\psi _0}{\prod_{\ell =1}^{N}{\sigma _{\ell}}}\sum_{m=0}^N{\frac{1}{m!}\left( \frac{2^{R_0}-1}{2^{R_0}\overline{\gamma }_{\mathrm{e}}} \right) ^m\sum_{n=0}^{K-1}{}}
\\
&\left( \begin{array}{c}
	K-1\\
	n\\
\end{array} \right) \left( -1 \right) ^n\left( \frac{1}{n+1} \right) ^{N-m+1}\left( 2^{R_0}\overline{\gamma }_{\mathrm{e}} \right) ^N.
\end{aligned}
\end{small}
\end{equation}

Since $\frac{1}{\left( n+1 \right) ^k}=\frac{1}{\left( k-1 \right) !}\int_0^1{t^n\left( -\ln t \right) ^{k-1}dt}$ holds, it follows that
\begin{equation}
\begin{small}
\label{eq98}
\begin{array}{l}
$$\sum_{n=0}^{K-1}{\left( \begin{array}{c}
	K-1\\
	n\\
\end{array} \right) \left( -1 \right) ^n\left( \frac{1}{n+1} \right) ^{N-m+1}}
\\
=\frac{1}{\left( N-m \right) !}\int_0^1{\left( -\ln t \right) ^{N-m}\left( 1-t \right) ^{K-1}dt}>0$$.
\end{array}
\end{small}
\end{equation}

Thus, the coefficient of $z^N$ is positive. Accordingly, in Bob's high-SNR region, (\ref{eq57}) can be rewritten as follows:
\begin{equation}
\begin{small}
\label{eq99}
\begin{array}{l}
\underset{\overline{\gamma }_{\mathrm{b}}\rightarrow \infty}{\lim}P_{\mathrm{SOP}_{\mathrm{M}}}\left( R_0 \right) \approx \left\{ \left[ \left( \prod_{\ell =1}^{\mathsf{DOF}}{\sigma _{\ell}} \right) /\left( K\psi _0 \right. \right. \right.
\\
\cdot\sum_{m=0}^{\mathsf{DOF}}{\frac{1}{m!}}\left( \frac{2^{R_0}-1}{2^{R_0}\overline{\gamma }_{\mathrm{e}}} \right) ^m\sum_{n=0}^{K-1}{\left( \begin{array}{c}
	K-1\\
	n\\
\end{array} \right)}\left( -1 \right) ^n
\\
\cdot\left. \left. \left. \left( \frac{1}{n+1} \right) ^{\mathsf{DOF}-m+1} \right) \right] ^{\frac{1}{\mathsf{DOF}}}\frac{\overline{\gamma }_{\mathrm{b}}}{2^{R_0}\overline{\gamma }_{\mathrm{e}}} \right\} ^{-\mathsf{DOF}}.
\end{array}
\end{small}
\end{equation}
Therefore, the diversity order is $\mathsf{DOF}$, and the array gain is given in (\ref{eq58}).
$\hfill\blacksquare$
\vspace{-0.4cm}
\section{Secrecy Rate Under Multiple Collaborative Eves}
Under multiple collaborative Eves, the secrecy rate is given in (\ref{eq100}) on the top of the next page.
\begin{figure*}
\begin{small}
\begin{equation}
\label{eq100}
\begin{array}{l}
R_{\mathrm{MC}}\approx \frac{1}{\overline{\gamma }_{\mathrm{b}}\overline{\gamma }_{\mathrm{e}}^{K}\Gamma \left( K \right) \prod_{\ell =1}^{\mathsf{DOF}}{\sigma _{b,\ell}}}\sum_{q=0}^{\infty}{\frac{\psi _q}{\sigma _{\min}^{q}\Gamma (\mathsf{DOF}+q)}\left\{ \left( \frac{1}{\overline{\gamma }_{\mathrm{b}}} \right) ^{\mathsf{DOF}+q-1}e^{\frac{1}{\overline{\gamma }_{\mathrm{b}}\sigma _{\min}}}\sum_{k=0}^{\mathsf{DOF}+q-1}{\left( \begin{array}{c}
	\mathsf{DOF}+q-1\\
	k\\
\end{array} \right) \left( -1 \right) ^{\mathsf{DOF}+q-1-k}} \right.}
\\
\cdot\left( \overline{\gamma }_{\mathrm{b}}\sigma _{\min} \right) ^{k+1}\left[ \frac{k!}{\ln 2}\left( e^{\frac{1}{\overline{\gamma }_{\mathrm{e}}}}\sum_{a=0}^{K-1}{\left( \begin{array}{c}
	K-1\\
	a\\
\end{array} \right) \left( -1 \right) ^{K-a-1}}\left( \frac{\overline{\gamma }_{\mathrm{b}}\overline{\gamma }_{\mathrm{e}}\sigma _{\min}}{\overline{\gamma }_{\mathrm{b}}\sigma _{\min}+\overline{\gamma }_{\mathrm{e}}} \right) ^{a+1}\left[ \mathrm{F}\left( a+1,\frac{\overline{\gamma }_{\mathrm{b}}\sigma _{\min}+\overline{\gamma }_{\mathrm{e}}}{\overline{\gamma }_{\mathrm{b}}\overline{\gamma }_{\mathrm{e}}\sigma _{\min}} \right) \right. \right. \right.
\\
\left. +\ln \frac{\overline{\gamma }_{\mathrm{e}}}{\overline{\gamma }_{\mathrm{b}}\sigma _{\min}+\overline{\gamma }_{\mathrm{e}}}\Gamma \left( a+1,\frac{\overline{\gamma }_{\mathrm{b}}\sigma _{\min}+\overline{\gamma }_{\mathrm{e}}}{\overline{\gamma }_{\mathrm{b}}\overline{\gamma }_{\mathrm{e}}\sigma _{\min}} \right) \right] +\left[ \left( \overline{\gamma }_{\mathrm{e}} \right) ^K\left( K-1 \right) !\left[ \mathrm{E}_1\left( \frac{1}{\overline{\gamma }_{\mathrm{b}}\sigma _{\min}} \right) -e^{\frac{1}{\overline{\gamma }_{\mathrm{e}}}}\mathrm{E}_1\left( \frac{\overline{\gamma }_{\mathrm{b}}\sigma _{\min}+\overline{\gamma }_{\mathrm{e}}}{\overline{\gamma }_{\mathrm{b}}\overline{\gamma }_{\mathrm{e}}\sigma _{\min}} \right) \right] \right.
\\
-\overline{\gamma }_{\mathrm{e}}e^{\frac{1}{\overline{\gamma }_{\mathrm{e}}}}\left[ \left( -1 \right) ^{K-1}\mathrm{E}_1\left( \frac{\overline{\gamma }_{\mathrm{b}}\sigma _{\min}+\overline{\gamma }_{\mathrm{e}}}{\overline{\gamma }_{\mathrm{b}}\overline{\gamma }_{\mathrm{e}}\sigma _{\min}} \right) +\sum_{t=1}^{K-1}{\left( \begin{array}{c}
	K-1\\
	t\\
\end{array} \right) \left( -1 \right) ^{K-1-t}\left( \frac{\overline{\gamma }_{\mathrm{e}}\overline{\gamma }_{\mathrm{b}}\sigma _{\min}}{\overline{\gamma }_{\mathrm{b}}\sigma _{\min}+\overline{\gamma }_{\mathrm{e}}} \right) ^t\Gamma \left( t,\frac{\overline{\gamma }_{\mathrm{b}}\sigma _{\min}+\overline{\gamma }_{\mathrm{e}}}{\overline{\gamma }_{\mathrm{e}}\overline{\gamma }_{\mathrm{b}}\sigma _{\min}} \right)} \right]
\\
-e^{\frac{1}{\overline{\gamma }_{\mathrm{e}}}}\sum_{w=2}^{K-1}{\left( \overline{\gamma }_{\mathrm{e}} \right) ^w}\left( \prod_{i=1}^{w-1}{\left( K-i \right)} \right) \left[ \left( -1 \right) ^{K-w}\mathrm{E}_1\left( \frac{\overline{\gamma }_{\mathrm{b}}\sigma _{\min}+\overline{\gamma }_{\mathrm{e}}}{\overline{\gamma }_{\mathrm{b}}\overline{\gamma }_{\mathrm{e}}\sigma _{\min}} \right) +\sum_{t=1}^{K-w}{\left( \begin{array}{c}
	K-w\\
	t\\
\end{array} \right) \left( -1 \right) ^{K-w-t}\left( \frac{\overline{\gamma }_{\mathrm{e}}\overline{\gamma }_{\mathrm{b}}\sigma _{\min}}{\overline{\gamma }_{\mathrm{b}}\sigma _{\min}+\overline{\gamma }_{\mathrm{e}}} \right) ^t} \right.
\\
\cdot\left. \left. \Gamma \left( t,\frac{\overline{\gamma }_{\mathrm{b}}\sigma _{\min}+\overline{\gamma }_{\mathrm{e}}}{\overline{\gamma }_{\mathrm{e}}\overline{\gamma }_{\mathrm{b}}\sigma _{\min}} \right) \right] \right] +\sum_{p=0}^{k-1}{\frac{1}{\left( k-p \right) !}}\left( e^{\frac{1}{\overline{\gamma }_{\mathrm{e}}}}\sum_{b=0}^{K-1}{\left( \begin{array}{c}
	K-1\\
	b\\
\end{array} \right) \left( -1 \right) ^{K-b-1}}\left( \frac{\overline{\gamma }_{\mathrm{e}}\overline{\gamma }_{\mathrm{b}}\sigma _{\min}}{\overline{\gamma }_{\mathrm{b}}\sigma _{\min}+\overline{\gamma }_{\mathrm{e}}} \right) ^{b+1}\left( \frac{\overline{\gamma }_{\mathrm{e}}}{\overline{\gamma }_{\mathrm{b}}\sigma _{\min}+\overline{\gamma }_{\mathrm{e}}} \right) ^{k-p} \right.
\\
\cdot\left( \mathrm{F}\left( k-p+b+1,\frac{\overline{\gamma }_{\mathrm{b}}\sigma _{\min}+\overline{\gamma }_{\mathrm{e}}}{\overline{\gamma }_{\mathrm{b}}\overline{\gamma }_{\mathrm{e}}\sigma _{\min}} \right) +\ln \frac{\overline{\gamma }_{\mathrm{e}}}{\overline{\gamma }_{\mathrm{b}}\sigma _{\min}+\overline{\gamma }_{\mathrm{e}}}\Gamma \left( k-p+b+1,\frac{\overline{\gamma }_{\mathrm{b}}\sigma _{\min}+\overline{\gamma }_{\mathrm{e}}}{\overline{\gamma }_{\mathrm{b}}\overline{\gamma }_{\mathrm{e}}\sigma _{\min}} \right) \right) +\left( k-p-1 \right) !
\\
\cdot\left. \left. \sum_{r=0}^{k-p-1}{\frac{1}{r!}e^{\frac{1}{\overline{\gamma }_{\mathrm{e}}}}\sum_{c=0}^{K-1}{\left( \begin{array}{c}
	K-1\\
	c\\
\end{array} \right) \left( -1 \right) ^{K-1-c}}\left( \frac{\overline{\gamma }_{\mathrm{e}}\overline{\gamma }_{\mathrm{b}}\sigma _{\min}}{\overline{\gamma }_{\mathrm{b}}\sigma _{\min}+\overline{\gamma }_{\mathrm{e}}} \right) ^{c+1}\left( \frac{\overline{\gamma }_{\mathrm{e}}}{\overline{\gamma }_{\mathrm{b}}\sigma _{\min}+\overline{\gamma }_{\mathrm{e}}} \right) ^r\Gamma \left( r+c+1,\frac{\overline{\gamma }_{\mathrm{b}}\sigma _{\min}+\overline{\gamma }_{\mathrm{e}}}{\overline{\gamma }_{\mathrm{b}}\overline{\gamma }_\mathrm{e}\sigma _{\min}} \right)} \right) \right)
\\
+\log _2\left( \overline{\gamma }_{\mathrm{b}}\sigma _{\min} \right) k!\sum_{n=0}^k{\frac{1}{n!}e^{\frac{1}{\overline{\gamma }_{\mathrm{e}}}}\sum_{d=0}^{K-1}{\left( \begin{array}{c}
	K-1\\
	d\\
\end{array} \right) \left( -1 \right) ^{K-1-d}}\left( \frac{\overline{\gamma }_{\mathrm{e}}\overline{\gamma }_{\mathrm{b}}\sigma _{\min}}{\overline{\gamma }_{\mathrm{b}}\sigma _{\min}+\overline{\gamma }_{\mathrm{e}}} \right) ^{d+1}\left( \frac{\overline{\gamma }_{\mathrm{e}}}{\overline{\gamma }_{\mathrm{b}}\sigma _{\min}+\overline{\gamma }_{\mathrm{e}}} \right) ^n}
\\
\cdot\left. \Gamma \left( n+d+1,\frac{\overline{\gamma }_{\mathrm{b}}\sigma _{b,\min}+\overline{\gamma }_{\mathrm{e}}}{\overline{\gamma }_{\mathrm{b}}\overline{\gamma }_{\mathrm{e}}\sigma _{\min}} \right) \right] -\overline{\gamma }_{\mathrm{b}}\sigma _{\min}^{\mathsf{DOF}+q}\left( \mathsf{DOF}+q-1 \right) !\sum_{k=0}^{\mathsf{DOF}+q-1}{\frac{1}{k!}}\left( \frac{1}{\overline{\gamma }_{\mathrm{b}}\sigma _{\min}} \right) ^ke^{\frac{\overline{\gamma }_{\mathrm{b}}\sigma _{\min}+\overline{\gamma }_{\mathrm{e}}}{\overline{\gamma }_{\mathrm{b}}\overline{\gamma }_{\mathrm{e}}\sigma _{\min}}}\sum_{f=0}^{K+k-1}{\left( \begin{array}{c}
	K+k-1\\
	f\\
\end{array} \right)}
\\
\cdot\left( -1 \right) ^{K+k-1-f}\left( \frac{\overline{\gamma }_{\mathrm{b}}\overline{\gamma }_{\mathrm{e}}\sigma _{\min}}{\overline{\gamma }_{\mathrm{b}}\sigma _{\min}+\overline{\gamma }_{\mathrm{e}}} \right) ^{f+1}\left. \frac{1}{\ln 2}\left( \mathrm{F}\left( f+1,\frac{\overline{\gamma }_{\mathrm{b}}\sigma _{\min}+\overline{\gamma }_{\mathrm{e}}}{\overline{\gamma }_{\mathrm{b}}\overline{\gamma }_{\mathrm{e}}\sigma _{\min}} \right) +\ln \left( \frac{\overline{\gamma }_{\mathrm{b}}\overline{\gamma }_{\mathrm{e}}\sigma _{\min}}{\overline{\gamma }_{\mathrm{b}}\sigma _{\min}+\overline{\gamma }_{\mathrm{e}}} \right) \Gamma \left( f+1,\frac{\overline{\gamma }_{\mathrm{b}}\sigma _{\min}+\overline{\gamma }_{\mathrm{e}}}{\overline{\gamma }_{\mathrm{b}}\overline{\gamma }_{\mathrm{e}}\sigma _{\min}} \right) \right) \right\}.
\end{array}
\end{equation}
\end{small}
\vspace{-0.8cm}
\end{figure*}
\vspace{-0.6cm}
\section{Diversity Order and Array Gain Under Multiple Collaborative Eves}
To make the expression concise, $1/\overline{\gamma }_{\mathrm{b}}$ and $\mathsf{DOF}$ are denoted as $z$ and $N$, respectively. Following a proof similar to Theorem 1, it can be observed that in Bob's high-SNR region, the coefficients of the terms $z^k$, for $k\in \mathbb{Z} ^+$ and $0\leqslant k\leqslant N-1$, in (\ref{eq69}) are zero. The coefficient of $z^N$ is
\begin{equation}
\begin{small}
\label{eq101}
\begin{aligned}
\mathrm{coeff}\left( z^N \right) =&\frac{\psi _0}{\prod_{\ell =1}^N{\sigma _{\ell}}}\sum_{m=0}^N{\frac{1}{m!}\left( \frac{2^{R_0}-1}{2^{R_0}\overline{\gamma }_{\mathrm{e}}} \right) ^m}
\\
&\cdot \left( \begin{array}{c}
	N-m+K-1\\
	K-1\\
\end{array} \right) \left( 2^{R_0}\overline{\gamma }_{\mathrm{e}} \right) ^N.
\end{aligned}
\end{small}
\end{equation}
Therefore, in Bob's high-SNR region, (\ref{eq69}) can be rewritten as follows:
\begin{equation}
\begin{small}
\label{eq102}
\begin{array}{l}
\underset{\overline{\gamma }_{\mathrm{b}}\rightarrow \infty}{\lim}P_{\mathrm{SOP}_{\mathrm{MC}}}\left( R_0 \right) \approx \left\{ \left[ \left( \prod_{\ell =1}^{\mathsf{DOF}}{\sigma _{\ell}} \right) /\left( \psi _0\sum_{m=0}^{\mathsf{DOF}}{\frac{1}{m!}} \right. \right. \right.
\\
\cdot\left. \left. \left. \left( \frac{2^{R_0}-1}{2^{R_0}\overline{\gamma }_{\mathrm{e}}} \right) ^m\left( \begin{array}{c}
	\mathsf{DOF}-m+K-1\\
	K-1\\
\end{array} \right) \right) \right] ^{\frac{1}{\mathsf{DOF}}}\frac{\overline{\gamma }_{\mathrm{b}}}{2^{R_0}\overline{\gamma }_{\mathrm{e}}} \right\} ^{-\mathsf{DOF}}.
\end{array}
\end{small}
\end{equation}
Thus, the corresponding diversity order is $\mathsf{DOF}$, and the array gain is given in (\ref{eq70}).
$\hfill\blacksquare$

\end{appendices}

\bibliographystyle{IEEEtran}
\vspace{-0.4cm}
\bibliography{references}

\end{document}